\documentclass[final]{svjour3}       
\usepackage[linesnumbered]{algorithm2e}
\usepackage{makecell}
\usepackage{array}
\usepackage{subfigure}
\usepackage{caption}
\usepackage{url}
\usepackage{multirow}
\usepackage{bm}
\usepackage{amsmath,amssymb}   
\usepackage{latexsym}   
\usepackage{graphicx}
\usepackage{setspace}
\usepackage{colortbl}  
\usepackage{acro}
\usepackage{lineno}
\usepackage[title]{appendix}

\usepackage{tikz}
\usetikzlibrary{calc,shapes,shadows,arrows}
\usetikzlibrary{decorations.pathmorphing} 
\usetikzlibrary{fit}					

\DeclareAcronym{ann}{short=ANN,long=Artificial Neural Network,
	class= abbrev}
\DeclareAcronym{fm}{short=FM,long=Fuzzy Model,
	class= abbrev}

\DeclareAcronym{pso}{short=PSO,long=Particle Swarm Optimization,
	class= abbrev}
\DeclareAcronym{cg}{short=CG,long=Conjugate Gradient,
	class= abbrev}
\DeclareAcronym{ll}{short=LL,long=Log-Likelihood,
	class= abbrev}
\DeclareAcronym{nll}{short=NLL,long=Negative value of Log-Likelihood,
	class= abbrev}  
\DeclareAcronym{map}{short=MAP,long=Maximizing A Posterior,
	class= abbrev}
\DeclareAcronym{mc}{short=MC,long=Monte Carlo,
	class= abbrev}
\DeclareAcronym{mcmc}{short=MCMC,long=Markov Chain Monte Carlo,
	class= abbrev}
\DeclareAcronym{gp}{short=GP,long=Gaussian Process,
	class= abbrev}

\DeclareAcronym{cep}{short=CEP,long=Certainty Equivalence Principle,
	class= abbrev}
\DeclareAcronym{uav}{short=UAV,long=Unmanned Aerial Vehicle,
	class= abbrev}
\DeclareAcronym{gmv}{short=GMV,long=Generalized Minimum Variance,
	class= abbrev}

\DeclareAcronym{mpc}{short=MPC,long=Model Predictive Control,
	class= abbrev}
\DeclareAcronym{nmpc}{short=NMPC,long= Nonlinear Model Predictive Control,
	class= abbrev}  
\DeclareAcronym{smpc}{short=SMPC,long= Stochastic Model Predictive Control,
	class= abbrev}
\DeclareAcronym{rmpc}{short=RMPC,long= Robust Model Predictive Control,
	class= abbrev}   

\DeclareAcronym{mrac}{short=MRAC,long=Model References Adaptive Control,
	class= abbrev}

\DeclareAcronym{mse}{short=MSE,long=Mean Squared Error,
	class= abbrev}
\DeclareAcronym{mae}{short=MAE,long=Mean Absolute Error,
	class= abbrev}
\DeclareAcronym{se}{short=SE,long=Standard Error,
	class= abbrev}
\DeclareAcronym{smse}{short=SMSE,long=Standardized Mean Squared Error,
	class= abbrev}

\DeclareAcronym{vtol}{short=VTOL,long=Vertical Take Off and Landing,
	class= abbrev}
\DeclareAcronym{dof}{short=DOF,long=Degree-of-Freedom,
	class= abbrev}

\DeclareAcronym{pid}{short=PID,long=Proportional-Integral-Derivative,
	class= abbrev}
\DeclareAcronym{lqr}{short=LQR,long=Linear-Quadratic Regulator,
	class= abbrev}
\DeclareAcronym{lmi}{short=LMI,long=Linear Matrix Inequality,
	class= abbrev}   

\DeclareAcronym{mfac}{short=MFAC,long=Model-Free Adaptive Control,
	class= abbrev}

\DeclareAcronym{dp}{short=DP,long=Dynamic Programming,
	class= abbrev}
\DeclareAcronym{adp}{short=ADP,long=Approximate Dynamic Programming,
	class= abbrev}

\DeclareAcronym{lp}{short=LP,long= Linear Programming,
	class= abbrev}  
\DeclareAcronym{nlp}{short=NLP,long= Nonlinear Programming,
	class= abbrev}  

\DeclareAcronym{kkt}{short= KKT,long= Karush-Kahn-Tucker,
	class= abbrev}

\DeclareAcronym{qp}{short= QP,long= Quadratic Programming,
	class= abbrev}   

\DeclareAcronym{sqp}{short= SQP,long= Sequential Quadratic Programming,
	class= abbrev}

\DeclareAcronym{fpsqp}{short= FP-SQP,long= Feasibility-Perturbed Sequential Quadratic Programming,
	class= abbrev}   

\DeclareAcronym{mfcq}{short= MFCQ,long= Mangasarian-Fromovitz Constraint Qualification,
	class= abbrev}   

\DeclareAcronym{licq}{short= LICQ,long= Linear Independence Constraint Qualification,
	class= abbrev}

\DeclareAcronym{iae}{short= IAE,long= Integral Absolute Error,
	class= abbrev}    

\DeclareAcronym{bfgs}{short= BFGS,long= Broyden-Fletcher-Goldfarb-Shanno,
	class= abbrev}    

\everymath{\displaystyle}
\newcommand{\colourred}{black}

\begin{document}
	
\setpagewiselinenumbers
\modulolinenumbers[5]

\title{Gaussian Process Model Predictive Control of An Unmanned Quadrotor}

\author{Gang~Cao \and Edmund M-K~Lai \and Fakhrul~Alam}

\institute{Gang~Cao \at
	School of Engineering and Advanced Technology, Massey University, Auckland, New Zealand
	\email{g.cao@massey.ac.nz}           
	\and
	Edmund M-K~Lai \at
	Department of Information Technology and Software Engineering, Auckland University of Technology, Auckland, New Zealand
	\and
	Fakhrul~Alam \at
	School of Engineering and Advanced Technology, Massey University, Auckland, New Zealand
}

\date{Received: date / Accepted: date}

\maketitle

\begin{abstract}
	
\textcolor{\colourred}{The~\ac{mpc} trajectory tracking problem of an unmanned quadrotor 
with input and output constraints is addressed.
In this article, the dynamic models of the quadrotor are obtained purely from operational data in the form of probabilistic \ac{gp} models.
This is different from conventional models obtained through Newtonian analysis.
A hierarchical control scheme is used to handle the trajectory tracking problem with the
translational subsystem in the outer loop and the rotational subsystem in the inner loop.
Constrained \ac{gp} based \ac{mpc} are formulated separately for both subsystems.
The resulting~\ac{mpc} problems are typically nonlinear and non-convex. 
We derived a \ac{gp} based local dynamical model that allows these optimization problems to
be relaxed to convex ones which
can be efficiently solved with a simple active-set algorithm.
The performance of the proposed approach is compared with an existing unconstrained~\ac{nmpc}.
Simulation results show that the two approaches exibit similar trajectory tracking performance.
However, our approach has the advantage of incorporating constraints on the control inputs.
In addition, our approach only requires $20\%$ of the computational time for \ac{nmpc}.}


\end{abstract}

\acresetall

\acresetall
\SetAlFnt{\small}

\newcommand{\newargmin}{\mathop{\mathrm{argmin}}} 
\newcommand{\newargmax}{\mathop{\mathrm{argmax}}} 
\newcommand{\scaledegree}{0.8}
\definecolor{mygray}{RGB}{194,204,208}

\newcommand*{\params}{\varTheta}
\newcommand*{\hyperparas}{\bm{\theta}}
\newcommand*{\bighyperparas}{\bm{\Theta}}
\newcommand*{\meanvalue}{\bm{\mu}}
\newcommand*{\varvalue}{\bm{\Sigma}}

\newcommand{\doublefigWidth}{0.475}
\newcommand{\multifigWidth}{0.475}
\newcommand{\tabcolsepwidth}{10pt}
\newcommand{\longtabcolsepwidth}{20pt}

\SetKwRepeat{Do}{do}{while}%
\newcommand{\nosemic}{\renewcommand{\@endalgocfline}{\relax}}
\newcommand{\dosemic}{\renewcommand{\@endalgocfline}{\algocf@endline}}
\newcommand{\pushline}{\Indp}
\newcommand{\popline}{\Indm\dosemic}
\let\oldnl\nl
\newcommand{\nonl}{\renewcommand{\nl}{\let\nl\oldnl}}
\renewcommand{\KwIn}{\textbf{Initialization}}

\newcommand{\tabincell}[2]{\begin{tabular}{@{}#1@{}}#2\end{tabular}}

\section{Introduction}
\label{sec:introduction}

The quadrotor helicopter (or quadrotor for short) is an aerial vehicle with vertical take-off and landing capabilities.
It has received a lot of interests recently due to its simplicity, maneuverability, and payload capabilities~\cite{A-Alexis-SwitchMPC-QuadrotorUAV-2011,A-Abdolhosseini-MPC-QuadrotorUAV-2013}.
It has been used in various military and civilian tasks~\cite{A-Metni-UAV-BridgeInspection-2007,IC-Doherty-UAV-HumanBodyDetection-2007}.

Trajectory tracking is one of the basic functions performed by a quadrotor in autonomous flight.
Designing a control system to perform this function is challenging because 
the quadrotor's dynamics are highly nonlinear and are subjected to random external disturbances.
Several control approaches have previously been investigated with varying degrees of success.
They include linear techniques such as~\ac{pid} and~\ac{lqr} control~\cite{IP-Bouabdallah-PID/LQ-UAV-2004},
as well as nonlinear techniques such as
sliding mode~\cite{IP-Madani-SlidingModeAndBackStepping-UAV-2007} and backstepping control~\cite{IP-Huang-AdaptiveTrackingUsingBackstepping-UAV-2010}. 
More recently, 
due to the conceptual simplicity,
\ac{mpc} techniques have been used
in~\cite{A-Raffo-HinfinityControl-UAV-2010,A-Alexis-SwitchMPC-QuadrotorUAV-2011} based on the linearised model and in~\cite{A-Abdolhosseini-MPC-QuadrotorUAV-2013} based on the nonlinear model.
Moreover, physical constraints on the system inputs and outputs, 
which is important for quadrotors, 
could easily be included as appropriate penalty terms in the cost function that is used to compute the optimal control.

\textcolor{\colourred}{The performance of~\ac{mpc}} is highly dependent on how accurately the model
describes the dynamics of the system being controlled.
Conventionally, \textcolor{\colourred}{dynamical models} are derived from first principles through Newton-Euler~\cite{A-Zuo-QuadrotorTrackingControl-2010} 
or Euler-Lagrange based formalisms~\cite{IP-Bouabdallah-PID/LQ-UAV-2004}.
Alternatively, empirical input-output data could be collected from a real, working quadrotor.
These data could then be used to construct a \ac{fm}~\cite{IP-Han-UAVfuzzymodelcontrol-2014} or an \ac{ann} model~\cite{IP-Voos-offlineNNbasedUAVcontrol-2007,A-Dierks-NNbasedUAVcontrol-2010}.
This data-driven approach has the advantage that unknown dynamics that are not considered by Newtonian analysis
could be captured by the empirical observations.
However, it is difficult to evaluate the quality of these \ac{fm} and \ac{ann} models.
\ac{gp} modelling is an alternative data-driven technique based on Bayesian theory.
Compared to~\ac{ann} and~\ac{fm}, 
a major advantage is that the quality of obtained~\ac{gp} model can be directly evaluated by~\ac{gp} variances which are naturally computed during the modelling and prediction processes.
\ac{gp} based technique has recently been used to learn the flight model of \ac{uav} \cite{IP-Hemakumara-IndentificationUAVusingDGP-2011,IP-Hemakumara-IndentificationUAVusingDGP-2013} and quadrotors \cite{IP-Berkenkamp-RboustLBNMPC-2014,IP-GPMPC4Quad-Gang-2016b}.

\textcolor{\colourred}{The cost functions used in early~\ac{gp} based~\ac{mpc} problems are deterministic even though
the \ac{gp} models are probabilistic \cite{IP-Kocijan-MPC-2003,IP-Kocijan-MPC-GP-2004,A-Likar-MPC-2007,IP-Grancharova-ApproxExplicitNMPC-2007}.
This issue has been addressed recently in~\cite{A-Klenske-GPMPC4Periodic-2015,IP-GPMPC4LTV-Gang-2016a,IP-GPMPC4Quad-Gang-2016b}
where the expectation of the cost function is used instead,
as proposed in~\cite{A-Mesbah-SMPC-2016}.
However, these works did not take into consideration any constraint on system inputs and outputs.
In addition, a computationally efficient method is required to solve the resulting
\ac{gp} based~\ac{mpc} optimization problem which is usually nonlinear and non-convex.}

\textcolor{\colourred}{In this article, a hierarchical control scheme is applied
to the trajectory tracking problem of a quadrotor,  
where a translational subsystem forms the outer loop and a rotational subsystem is in the inner loop
\cite{A-Raffo-HinfinityControl-UAV-2010,A-Alexis-SwitchMPC-QuadrotorUAV-2011}.
Each subsystem is independently modelled by a~\ac{gp} model.
We propose a \ac{gp} based \ac{mpc} control scheme, referred to as GPMPC, solve the resulting two~\ac{mpc} tracking problems.
It tackles the issues mentioned above regarding the objective function and computational efficiency.
The performance of GPMPC is evaluated by simulations on two non-trivial trajectories.}

\section{Quadrotor System Modelling Using GP}
\label{sec:gpmodelling}

The quadrotor can be viewed as a 6~\ac{dof} rigid body with generalized coordinates $\mathbf{q}=[x,y,z,\phi,\theta,\psi]^T\in\mathbb{R}^6$,
\textcolor{\colourred}{where $x,y,z$ denotes the quadrotor's positions w.r.t. earth-fixed frame (E-frame) 
and $\phi,\theta,\psi$ represents quadrotor's attitudes w.r.t. body-fixed frame (B-Frame).
Motion is controlled by a main thrust $U_1$ and three torques $U_2$, $U_3$ and $U_4$}.
Thus it is an underactuated system.
Furthermore, \textcolor{\colourred}{the dynamical model of the quadrotor is defined by the state-space function
$\ddot{\mathbf{q}}=f_\mathbf{q}(\mathbf{q},\dot{\mathbf{q}}, U_1, U_2, U_3, U_4)$ 
which is usually nonlinear~\cite{A-Raffo-HinfinityControl-UAV-2010}.}
In order to simplify the control of the quadrotor, the system is typically decomposed into 
two subsystems -- a translational subsystem and a rotational subsystem.
Let the system state of the translational subsystem be $\mathbf{x}^\xi=[x,\dot{x},y,\dot{y},z,\dot{z}]^T\in\mathbb{R}^6$ 
and its control be $\mathbf{u}^\xi=[U_1,u_x,u_y]^T\in\mathbb{R}^3$.
The dynamics of this subsystem can be described by~\textcolor{red}{\cite{A-Raffo-HinfinityControl-UAV-2010}}
\begin{equation}\label{eqn:translationSSfunc}
\dot{\mathbf{x}}^\xi= f_\xi\left(\mathbf{x}^\xi, \mathbf{u}^\xi\right)+ \bm{\epsilon}^\xi
\end{equation}
where $f_\xi:\mathbb{R}^6\times\mathbb{R}^3\rightarrow\mathbb{R}^6$ is nonlinear and is usually corrupted by white noises $\bm{\epsilon}^\xi\in\mathbb{R}^6$.
$u_x$ and $u_y$ are two intermediate controls to actuate the translational subsystem and are given by
\begin{equation}\label{eqn:intercontrols}
\begin{aligned}
u_x&=\cos{\phi}\sin{\theta}\cos{\psi}+\sin{\phi}\sin{\psi}\\
u_y&=\cos{\phi}\sin{\theta}\sin{\psi}-\sin{\phi}\cos{\psi}\\
\end{aligned}
\end{equation}

Similarly, let
$\mathbf{x}^\eta = [\phi,\dot{\phi},\theta,\dot{\theta},\psi,\dot{\psi}]^T\in\mathbb{R}^6$ and  $\mathbf{u}^\eta=[U_2,U_3,U_4]^T\in\mathbb{R}^3$ be the state and control for
the rotational subsystem.
Its system equation is given by~\textcolor{red}{\cite{A-Raffo-HinfinityControl-UAV-2010}}
\begin{equation}\label{eqn:rotationSSfunc}
\dot{\mathbf{x}}^\eta= f_\eta\left(\mathbf{x}^\eta, \mathbf{u}^\eta\right)+ \bm{\epsilon}^\eta
\end{equation}
where $f_\eta:\mathbb{R}^6\times\mathbb{R}^3\rightarrow\mathbb{R}^6$ is another nonlinear function and $\bm{\epsilon}^\eta\in\mathbb{R}^6$ represents the white noise.

\subsection{GP Modelling}
\label{subsec:gpmodelling}

\textcolor{\colourred}{The system equations (\ref{eqn:translationSSfunc}) and (\ref{eqn:rotationSSfunc}) of both subsystems
can be expressed in the following general form in the discrete-time domain by
\begin{equation}
\mathbf{x}_{k+1} = f(\mathbf{x}_k, \mathbf{u}_k) +\mathbf{w}_k
\label{eqn:nltvsys}
\end{equation}
where $\mathbf{x}_k\in\mathbb{R}^n$ denotes an $n$-dimensional state vector and $\mathbf{u}_k\in\mathbb{R}^m$ 
represents an $m$-dimensional input vector at the sampling time $k$.
$f:\mathbb{R}^{n}\times\mathbb{R}^{m}\rightarrow\mathbb{R}^{n}$ is a discrete nonlinear function, and
$\mathbf{w}_k\in~\mathbb{R}^n$ is Gaussian white noise.
To learn such an unknown function $f(\cdot)$ using \ac{gp} modelling techniques,
a natural choice for the model inputs and outputs are the state-control tuple 
$\tilde{\mathbf{x}}_k = (\mathbf{x}_k,\mathbf{u}_k)\in\mathbb{R}^{n+m}$ and the next state $\mathbf{x}_{k+1}$ respectively.
However, in practice, the difference $\Delta\mathbf{x}_k=\mathbf{x}_{k+1}-\mathbf{x}_{k}\in\mathbb{R}^n$ 
is usually smaller less than the values of $\mathbf{x}_k$.
Thus it is more advantageous to use $\Delta\mathbf{x}_k$ as the model output instead~\cite{PHD-Deisenroth-EfficientRLusingGP-2010}}.

A~\ac{gp} model is completely specified by its mean and covariance function~\cite{B-GPMLbook-2006}.
Assuming that the mean of the model input $\tilde{\mathbf{x}}_k$ is zero, the squared exponential covariance is given by $\mathbf{K}(\tilde{\mathbf{x}}_i,\tilde{\mathbf{x}}_j)=\sigma_s^2 \exp(-\frac{1}{2}(\tilde{\mathbf{x}}_i-\tilde{\mathbf{x}}_j)^T\bm{\varLambda}(\tilde{\mathbf{x}}_i-\tilde{\mathbf{x}}_j))+\sigma_n^2$, where $i$ and $j$ denote two sampling time steps.
The parameters
$\sigma_s^2,\sigma_n^2$ and the entries of matrix 
\textcolor{\colourred}{$\bm{\varLambda}\in\mathbb{R}^{(n+m)\times(n+m)}$ (usually is a diagonal matrix)} are referred to as the 
hyperparameters $\hyperparas$ of a~\ac{gp} model.
Given $D$ training inputs \textcolor{\colourred}{$\tilde{\mathbf{X}}=[\tilde{\mathbf{x}}_1,\cdots,\tilde{\mathbf{x}}_D]\in\mathbb{R}^{(n+m)\times D}$} and their corresponding training targets \textcolor{\colourred}{$\mathbf{y}=[\Delta\mathbf{x}_1,\cdots,\Delta\mathbf{x}_D]^T\in\mathbb{R}^{nD}$},
the joint distribution between $\mathbf{y}$ and a test target $\Delta\mathbf{x}^*_k$ \textcolor{\colourred}{corresponding to the test input $\tilde{\mathbf{x}}_k^*$ at sampling time $k$} is assumed to follow a Gaussian distribution. That is
\begin{equation}
p\Bigg(\begin{array}{c} \mathbf{y}\\ \Delta\mathbf{x}^*_k
\end{array}\Bigg)
\sim \mathcal{N}\Bigg( 
\textcolor{\colourred}{\mathbf{0}}, \begin{array}{cc}
\mathbf{K}(\tilde{\mathbf{X}}, \tilde{\mathbf{X}})+\sigma_n\mathbf{I} & \mathbf{K}(\tilde{\mathbf{X}}, \tilde{\mathbf{x}}^*_k)\\
\mathbf{K}(\tilde{\mathbf{x}}^*_k, \tilde{\mathbf{X}}) & \mathbf{K}(\tilde{\mathbf{x}}^*_k, \tilde{\mathbf{x}}^*_k)
\end{array}
\Bigg)
\end{equation}
\textcolor{\colourred}{where $\mathcal{N}(\cdot)$ denotes a multivariate Gaussian distribution and $\mathbf{0}\in\mathbb{R}^{nD}$} is a zero vector.
In addition,
the posterior distribution over the observations can be obtained by restricting the joint distribution to only contain those targets that agree with the observations.
This is achieved by conditioning the joint distribution on the observations,
and results in the predictive mean and variance function as follows~\cite{B-GPMLbook-2006}
\begin{subequations}\label{eqn:meanvar}
	\begin{align}
	m(\tilde{\mathbf{x}}^*_k)=\textit{E}_f[\Delta\mathbf{x}^*_k]&=\mathbf{K}(\tilde{\mathbf{x}}^*_k, \tilde{\mathbf{X}})\mathbf{K}_\sigma^{-1}\mathbf{y}\\
	\sigma^2(\tilde{\mathbf{x}}^*_k)=\textit{Var}_f[\Delta\mathbf{x}^*_k] &=
	\mathbf{K}(\tilde{\mathbf{x}}^*_k, \tilde{\mathbf{x}}^*_k)\\
	\nonumber&-\mathbf{K}(\tilde{\mathbf{x}}^*_k, \tilde{\mathbf{X}})\mathbf{K}_\sigma^{-1}\mathbf{K}(\tilde{\mathbf{X}}, \tilde{\mathbf{x}}^*_k)
	\end{align}
\end{subequations}
where $\mathbf{K}_\sigma=\mathbf{K}(\tilde{\mathbf{X}}, \tilde{\mathbf{X}})+\sigma_n\mathbf{I}$.
The state \textcolor{\colourred}{at the next sampling time $k+1$} also follows a Gaussian distribution. That is
\begin{equation}
\label{eqn:nextstate}
p(\mathbf{x}_{k+1})\sim\mathcal{N}(\meanvalue_{k+1},\varvalue_{k+1})
\end{equation} 
where
\begin{subequations}\label{eqn:statedistribute-1}
	\begin{align}
	\meanvalue_{k+1}&=\mathbf{x}_k+m(\tilde{\mathbf{x}}^*_k)\\
	\quad \varvalue_{k+1}&=\sigma^2(\tilde{\mathbf{x}}^*_k)
	\end{align}
\end{subequations}

Typically, the hyperparameters of the \ac{gp} model are learned by maximizing the log-likelihood function given by
\begin{equation}
\begin{aligned}
\log p(\mathbf{y}|\tilde{\mathbf{X}},\hyperparas)=&
-\frac{1}{2}\mathbf{y}^T\mathbf{K}_\sigma^{-1}\mathbf{y}
-\frac{1}{2}\log\left|\mathbf{K}_\sigma^{-1}\right|\\
&-\frac{D}{2}\log(2\pi)
\end{aligned}
\label{eqn:loglikelihood}
\end{equation}
This results in a nonlinear non-convex optimization problem that is traditionally solved by using~\ac{cg} or~\ac{bfgs} algorithms.

\subsection{Uncertainty propagation}
\label{subsec:undertaintypropagate}

With the \ac{gp} model obtained, one-step-ahead predictions can be made by using (\ref{eqn:meanvar}) 
and (\ref{eqn:statedistribute-1}).
When multiple-step predictions are required,
the conventional way is to iteratively perform multiple one-step-ahead predictions using the estimated mean values.
However, this process does not take into account the uncertainties introduced by each successive prediction.
This issue has been shown to be important in time-series predictions~\cite{IP-Girard-MultiStepTimeSeriesForecasting-2003}.

The uncertainty propagation problem can be dealt with by assuming that the joint distribution of 
the training inputs is uncertain and follows a Gaussian distribution.
That is,
\begin{equation}
p(\tilde{\mathbf{x}}_k)=p\big(\mathbf{x}_k,\mathbf{u}_k\big)\sim \mathcal{N}(\tilde{\meanvalue}_k,\tilde{\varvalue}_k)
\end{equation} 
with mean and variance given by
\begin{subequations}\label{eqn:augementedstateMeanVar}
	\begin{align}
	\tilde{\meanvalue}_k &=\left[\meanvalue_k,\textit{E}\left[\mathbf{u}_k\right]\right]^T \\
	\tilde{\varvalue}_k &=\left[\begin{array}{cc}
	\varvalue_k & \textit{Cov}\left[\mathbf{x}_k,\mathbf{u}_k\right]\\
	\textit{Cov}\left[\mathbf{u}_k,\mathbf{x}_k\right]&
	\textit{Var}\left[\mathbf{u}_k\right]
	\end{array}
	\right]
	\end{align}
\end{subequations}
where 
$\textit{Cov}\left[\mathbf{x}_k,\mathbf{u}_k\right] =
\textit{E}\left[\mathbf{x}_k\mathbf{u}_k\right]-\meanvalue_k\textit{E}\left[\mathbf{u}_k\right]$. 
Here, $\textit{E}\left[\mathbf{u}_k\right]$ and $\textit{Var}\left[\mathbf{u}_k\right]$ are the mean and variance of the system controls.

The exact predictive distribution of the training target could then be obtained by integrating over the training input distribution: 
\begin{equation}
p(\Delta\mathbf{x}^*_k)=\int p(f(\tilde{\mathbf{x}}^*_k)|\tilde{\mathbf{x}}^*_k) p(\tilde{\mathbf{x}}^*_k)d\tilde{\mathbf{x}}^*_k
\end{equation}
However, this integral is analytically intractable.
Numerical solutions can be obtained using Monte-Carlo simulation techniques.
In~\cite{IP-Candela-GPUncertainPropagation-2003}, a moment-matching based approach is proposed to obtain an analytical Gaussian approximation.
The mean and variance at an uncertain input can be obtained
through the laws of iterated expectations and conditional variances respectively~\cite{PHD-Deisenroth-EfficientRLusingGP-2010}.
They are given by
\begin{subequations}\label{eqn:pgpmeanvar}
	\begin{align}
	m(\tilde{\mathbf{x}}^*_k) & =
	\textit{E}_{\tilde{\mathbf{x}}^*_k}\Big[\textit{E}_f\big[\Delta\mathbf{x}^*_k\big]\Big]\\
	\sigma^2(\tilde{\mathbf{x}}^*_k) &=\textit{E}_{\tilde{\mathbf{x}}^*_k}\Big[\textit{Var}_f\big[\Delta\mathbf{x}^*_k\big]\Big]+\textit{Var}_{\tilde{\mathbf{x}}^*_k}\Big[\textit{E}_f\big[\Delta\mathbf{x}^*_k\big]\Big]
	\end{align}
\end{subequations}
Equation (\ref{eqn:statedistribute-1}) then becomes
\begin{subequations}\label{eqn:statedistribute-2}
	\begin{align}
	\meanvalue_{k+1} =& \meanvalue_k + m(\tilde{\mathbf{x}}^*_k)\\
	\varvalue_{k+1} =& \varvalue_k+ \sigma^2(\tilde{\mathbf{x}}^*_k)\\
	\nonumber&+\textit{Cov}\big[\mathbf{x}_k, \Delta\mathbf{x}_k\big]+\textit{Cov}\big[\Delta\mathbf{x}_k, \mathbf{x}_k\big]
	\end{align}
\end{subequations}

The computational complexity of~\ac{gp} inference using (\ref{eqn:pgpmeanvar}) is $\mathcal{O}(D^2n^2(n+m))$
which is quite high.
Hence, \ac{gp} is normally only suitable for problems with limited dimensions (under 12 as suggested by most publications) and limited size of training data.
For problems with higher dimensions,
sparse~\ac{gp} approaches~\cite{A-Sparse-Appro-2005} are often used.

\section{Control Problem Formulation}
\label{sec:probform}

\begin{figure*}[!t]
	\centering
	\tikzstyle{block}=[draw,rectangle,thick,minimum height=4em,
	minimum width=1em]
	\scalebox{0.65}{
		\begin{tikzpicture}[auto, node distance=2cm,>=latex']
		\node [block,align=center](trackgenerator){Trajectory \\Generator};
		\node [block, xshift=1.5cm,right of =trackgenerator,   align=center](transControl){Translation\\Controller};
		\node [block,right of=transControl,xshift=1.5cm,align=center](rotateControl){Rotation\\Controller};
		\node [block,right of=rotateControl,xshift=1.5cm,align=center](rotateSYS){Rotation\\ Subsystem};
		\node [block,right of=rotateSYS,xshift=2cm,yshift=1cm,align=center](transSYS){Translation\\ Subsystem}; 
		\draw[dashed, line width=2pt, fill=gray!50, fill opacity=0.2] ([xshift=1cm] rotateControl.east) --++(0,2.5)node[xshift=1cm,yshift=-0.5cm, text opacity=1 ]{Quadrotor} --++(6.75,0)--++(0,-3.5)--++(-6.75,0)--([xshift=1cm] rotateControl.east);
		\draw [->] (trackgenerator.east)node[xshift=0.75cm,yshift=1cm]{$\left[ \begin{array}{c}x_d\\y_d\\z_d\end{array}\right]$} -- (transControl);
		\draw [<-] ([yshift=-0.5cm] rotateControl.west)node[xshift=-0.75cm,yshift=-0.5cm]{$\psi_d = 0$}--++(-1,0);
		\draw [->] (transControl.east)node[xshift=0.75cm,yshift=0.5cm]{$\phi_d,\theta_d$}-- (rotateControl);
		\draw [->] ([yshift=0.5cm] rotateControl.east)node[xshift=0.5cm,yshift=0.25cm]{$U_2$} -- ([yshift=0.5cm] rotateSYS.west);
		\draw [->] (rotateControl.east)node[xshift=0.5cm,yshift=0.25cm]{$U_3$} -- (rotateSYS);
		\draw [->] ([yshift=-0.5cm] rotateControl.east)node[xshift=0.5cm,yshift=0.25cm]{$U_4$} -- ([yshift=-0.5cm] rotateSYS.west);
		\draw [->] ([xshift=0.75cm] transControl.north)node[xshift=4.1cm,yshift=0.75cm]{$U_1$}|-([yshift=3cm] transSYS);
		\draw [->] (rotateSYS) --++ (7,0)node{};
		\draw [->] ([xshift=1cm] rotateSYS.east) |- ([yshift=-1cm] transSYS);
		\draw [->] (transSYS.east) --++(2,0);
		\draw [->] ([xshift=0.5cm] transSYS.east)node[xshift=1cm,yshift=0.5cm]{$[x,y,z]^T$} --++(0,2.5)-|(transControl.north);
		\draw [->] ([xshift=4.5cm] rotateSYS.east)node[xshift=1cm,yshift=-0.5cm]{$[\phi,\theta,\psi]^T$} --++(0,-2.5)-|(rotateControl.south);
		\draw[<-, line width=2pt] (rotateSYS.south) --++(0,-1)node[yshift=-0.25cm]{Disturbance:$\bm{\epsilon}^\eta$};
		\draw[<-, line width=2pt] (transSYS.south) --++(0,-2)node[yshift=-0.25cm]{Disturbance:$\bm{\epsilon}^\xi$};
		\end{tikzpicture}}                            
	\caption{The Overall Control Scheme for Quadrotor}     
	\label{fig:overallstrategy}            
\end{figure*}

\subsection{MPC Problem for Subsystems}
With the quadrotor system decomposed into two subsystems,
a hierarchical structure as shown in Figure~\ref{fig:overallstrategy} can be used for the controller~\cite{A-Raffo-HinfinityControl-UAV-2010,A-Alexis-SwitchMPC-QuadrotorUAV-2011}.
In the outer loop,
the translational subsystem is controlled to follow a sequence of desired positions $[x_d,y_d,z_d]^T$.
The optimal control $U_1$ and two intermediate controls $u_x$ and $u_y$ are obtained 
by minimizing the tracking errors.
With $\psi_d =0$,
the desired attitudes $\theta_d$ and $\phi_d$ can be obtained using (\ref{eqn:intercontrols}).
Then, the rotational subsystem's attitudes $[\phi,\theta,\psi]^T$ are tuned to achieve 
the given target values in the inner loop.
By minimizing attitude errors,
the optimal controls $U_2$, $U_3$ and $U_4$ can be obtained.
Finally, the optimal control inputs $U_1$, $U_2$, $U_3$ and $U_4$ are applied to the quadrotor.

For a horizon of $H\geq 1$,
the discrete~\ac{mpc} trajectory tracking problem in the outer loop is given by
\begin{subequations}\label{eqn:trans-optproblem}
	\begin{align}
	\min_{\mathbf{u}^\xi(\cdot)}&\sum_{i=1}^{H}
	\bigg\{\left\|\mathbf{x}^\xi_{k+i}-\mathbf{r}^\xi_{k+i}\right\|_{\mathbf{Q}^\xi}^2
	+\left\|\mathbf{u}^{\xi}_{k+i-1}\right\|_{\mathbf{R}^\xi}^2\bigg\}\\
	\mbox{s.t.}\quad&\mathbf{x}^\xi_{k+i+1}=f_1(\mathbf{x}^\xi_{k+i},\mathbf{u}^\xi_{k+i-1})\\
	&\mathbf{x}^\xi_{\text{min}}\leq\mathbf{x}^\xi_{k+i}\leq\mathbf{x}^\xi_{\text{max}}\\
	&\mathbf{u}^\xi_{\text{min}}\leq\mathbf{u}^\xi_{k+i-1}\leq\mathbf{u}^\xi_{\text{max}}
	\end{align}
\end{subequations}
where the $f_1(\cdot)$ represents the \ac{gp} model of the translational subsystem.
$\big\|\cdot\big\|_{\mathbf{Q}^\xi}$ and $\big\|\cdot\big\|_{\mathbf{R}^\xi}$ denote the two $2$-norms weighted by positive definite matrices $\mathbf{Q}^\xi$ and $\mathbf{R}^\xi$ respectively.
$\mathbf{x}^\xi_{k+i}$ and $\mathbf{u}^{\xi}_{k+i-1}$ are the system states and control inputs, and
$\mathbf{r}^\xi_{k+i}=[x_{d,k+i}, y_{d,k+i}, z_{d,k+i}]^T$ denotes the desired positions at time $k+i$.
In addition,
$\mathbf{x}^{\xi}_{\text{max}}\geq\mathbf{x}^{\xi}_{\text{min}}$ and $\mathbf{u}^{\xi}_{\text{max}}\geq\mathbf{u}^{\xi}_{\text{max}}$ are the upper and lower bounds of the system states and control inputs respectively.

In the same way, for the inner loop,
the discrete~\ac{mpc} optimization problem is given by
\begin{subequations}\label{eqn:rotate-optproblem}
	\begin{align}
	\min_{\mathbf{u}^\eta(\cdot)}&\sum_{i=1}^{H}
	\bigg\{\left\|\mathbf{x}^\eta_{k+i}-\mathbf{r}^\eta_{k+i}\right\|_{\mathbf{Q}^\eta}^2
	+\left\|\mathbf{u}^\eta_{k+i-1}\right\|_{\mathbf{R}^\eta}^2\bigg\}\\
	\mbox{s.t.}\quad&\mathbf{x}^\eta_{k+i+1}=f_2(\mathbf{x}^\eta_{k+i},\mathbf{u}^\eta_{k+i-1})\\
	&\mathbf{x}^\eta_{\text{min}}\leq\mathbf{x}^\eta_{k+i}\leq\mathbf{x}^\eta_{\text{max}}\\
	&\mathbf{u}^\eta_{\text{min}}\leq\mathbf{u}^\eta_{k+i-1}\leq\mathbf{u}^\eta_{\text{max}}
	\end{align}
\end{subequations}
where $f_2(\cdot)$ represents the~\ac{gp} model of the rotational subsystem.

Problems (\ref{eqn:trans-optproblem}) and (\ref{eqn:rotate-optproblem}) can be
rewritten in the following general form:
\begin{subequations}\label{eqn:optproblem1}
	\begin{align}
	\mathbf{V}_k^*&=\min_{\mathbf{u}(\cdot)}\mathcal{J}(\mathbf{x}_k,\mathbf{u}_{k-1}, \mathbf{r}_k)\\
	\mbox{s.t.}\; & \mathbf{x}_{k+i|k}=f(\mathbf{x}_{k+i-1|k}, \mathbf{u}_{k+i-1})\\
	\label{eqn:constraints-x}&\mathbf{x}_{\text{min}} \leq \mathbf{x}_{k+i|k} \leq \mathbf{x}_{\text{max}}\\
	\label{eqn:constraints-u}&\mathbf{u}_{\text{min}} \leq\mathbf{u}_{k+i-1} \leq \mathbf{u}_{\text{max}}\\
	\nonumber& i=1,\cdots, H
	\end{align}
\end{subequations}
with the quadratic cost function
\begin{equation}
\begin{aligned}
&\mathcal{J}(\mathbf{x}_k,\mathbf{u}_{k-1},\mathbf{r}_k)\\
&=\sum_{i=1}^{H}\Big\{\big\|\mathbf{x}_{k+i}-\mathbf{r}_{k+i}\big\|^2_{\mathbf{Q}}+\big\|\mathbf{u}_{k+i-1}\big\|^2_{\mathbf{R}}\Big\}
\end{aligned}
\label{eqn:costfunction1}
\end{equation}
It should be noted that the control horizon is assumed to be equal to the prediction horizon $H$ in this paper.
In the rest of this article, the cost function $\mathcal{J}(\mathbf{x}_k,\mathbf{u}_{k-1},\mathbf{r}_k)$ shall be rewritten as $\mathcal{J}(\mathbf{x}_k,\mathbf{u}_{k-1})$ for brevity.

\subsection{MPC with GP Models}

When the dynamical system is described by a \ac{gp} model,
the original problem (\ref{eqn:optproblem1}) becomes a stochastic one~\cite{A-Grancharova-ExplicitMPC-2008}.
The minimization should be performed over the expected value of $\mathcal{J}(\cdot)$ instead and
the constraints are modified as follows.
\begin{subequations}\label{eqn:optproblem2}
	\begin{align}
	\mathbf{V}_k^\ast & =
	\min_{\mathbf{u}(\cdot)}\textit{E}\big[ \mathcal{J}(\mathbf{x}_k,\mathbf{u}_{k-1})\big]\\
	\mbox{s.t.}\quad & p(\mathbf{x}_{k+1}|\mathbf{x}_k)\sim\mathcal{N}(\meanvalue_{k+1},\varvalue_{k+1})\\
	&\mathbf{u}_{\text{min}} \leq\mathbf{u}_{k+i-1} \leq \mathbf{u}_{\text{max}}\\
	& p\big\{\mathbf{x}_{k+i|k}\geq \mathbf{x}_{\text{min}}\big\} \geq \eta \label{subeqn:chance1}\\
	& p\big\{\mathbf{x}_{k+i|k}\leq \mathbf{x}_{\text{max}}\big\} \geq \eta \label{subeqn:chance2}
	\end{align}
\end{subequations}
where $\eta$ denotes a confidence level. 
For $\eta = 0.95$, the chance constraints (\ref{subeqn:chance1})  and (\ref{subeqn:chance2})
are equivalent to
\begin{subequations}\label{eqn:deterministicConstraints}
	\begin{align}
	\meanvalue_{k+i}-2\varvalue_{k+i} &\geq \mathbf{x}_{\text{min}}\\
	\meanvalue_{k+i}+2\varvalue_{k+i} &\leq \mathbf{x}_{\text{max}}
	\end{align}
\end{subequations}
Given (\ref{eqn:costfunction1}),
\begin{equation}
\begin{aligned}
&\textit{E}\big[\mathcal{J}(\mathbf{x}_k,\mathbf{u}_{k-1})\big] \\
&=\textit{E}\Big[
\sum_{i=1}^{H}\Big\{\big\|\mathbf{x}_{k+i}-\mathbf{r}_{k+i}\big\|^2_{\mathbf{Q}}+\big\|\mathbf{u}_{k+i-1}\big\|^2_{\mathbf{R}}\} \Big] \\
&= \sum_{i=1}^{H}\textit{E}\Big[ \big\|\mathbf{x}_{k+i}-\mathbf{r}_{k+i}\big\|^2_{\mathbf{Q}}+\big\|\mathbf{u}_{k+i-1}\big\|^2_{\mathbf{R}}\Big] \\
&= \sum_{i=1}^{H}\bigg\{\textit{E}\Big[ \big\|\mathbf{x}_{k+i}-\mathbf{r}_{k+i}\big\|^2_{\mathbf{Q}}\Big]+\textit{E}\Big[\big\|\mathbf{u}_{k+i-1}\big\|^2_{\mathbf{R}}\Big]\bigg\}
\end{aligned}
\label{eqn:costfunction2-1}
\end{equation}
In practice, the controls are deterministic. 
Hence, $\textit{E}\big[\mathbf{u}_k^2\big]=\mathbf{u}_k^2$ 
and (\ref{eqn:costfunction2-1}) becomes
\begin{equation}
\begin{aligned}
&\textit{E}\big[\mathcal{J}(\mathbf{x}_k,\mathbf{u}_{k-1})\big]\\
&=\sum_{i=1}^{H}\Big\{\textit{E}\bigg[\big\|\mathbf{x}_{k+i}-\mathbf{r}_{k+i}\big\|^2_{\mathbf{Q}}\bigg]+\big\|\mathbf{u}_{k+i-1}\big\|^2_{\mathbf{R}}\Big\}\\
&=\sum_{i=1}^{H}\Big\{\big\|\meanvalue_{k+i}-\mathbf{r}_{k+i}\big\|^2_{\mathbf{Q}}+\big\|\mathbf{u}_{k+i-1}\big\|^2_{\mathbf{R}}+\textit{trace}\big(\mathbf{Q}\varvalue_{k+i}\big)\Big\} \\
&=h \left(\meanvalue_{k}, \mathbf{u}_{k-1} \right)
\end{aligned}
\label{eqn:costfunction2-2}
\end{equation}
\textcolor{\colourred}{The elaboration of (\ref{eqn:costfunction2-2}) can be found in Appendix~\ref{sec:appendix}}.
With this cost function and the state constraints (\ref{eqn:deterministicConstraints}), 
we are able to relax the original stochastic optimization problem (\ref{eqn:optproblem2}) to a deterministic nonlinear one.
Furthermore, the resulting deterministic cost function involves the model variance $\varvalue$.
This allows model uncertainties to be explicitly included in the computation of optimized controls.

\section{Proposed Solution}
\label{sec:gpmpc}

\textcolor{\colourred}{Solving the constrained~\ac{mpc} optimization problem (\ref{eqn:costfunction2-2}) with state constraints (\ref{eqn:deterministicConstraints})
	is not simple because it is typically nonlinear and non-convex.
	Solving non-convex problems due to they are computationally complicated and have multiple local optima.
	This significantly limits the application of~\ac{mpc} in real world problems.
	An effective and efficient solution method is therefore very important \cite{Valipour-JIRS-2014,A-Yannopoulos-JIRS-2015,Valipour-JIRS-2016-1,Valipour-JIRS-2016-2,Valipour-JIRS-2016-3,Valipour-JIRS-2017}.
	A conventional approach is to use derivative-based methods such as~\ac{sqp} and interior-point algorithms~\cite{IP-Diehl-EfficientSolution4MPC-2009}.
	When the derivatives of the cost function are unavailable or are too difficult to compute,
	they could be iteratively approximated by using sampling methods~\cite{A-Lucidi-DeviativeFreeOpt-2002,A-Liuzzi-SequentialDerivativeFreeOpt-2010}.
	An alternative solution is to use evolutionary algorithms such as~\ac{pso}~\cite{A-Luo-PSO4NLP-2007}.
	A more complete review of solution methods can be found in~\cite{IP-Diehl-EfficientSolution4MPC-2009}}.

\textcolor{\colourred}{In this section, we present our proposed solution which is by local linearization.
This allows the original problem to be relaxed into a convex one which can then be solved efficiently by
active-set methods.}

\subsection{GP Based Local Dynamical Model}
\label{subsec:gplocalmodel}

There are many different ways by which a \ac{gp} model could be linearised.
In~\cite{IP-Berkenkamp-RboustLBNMPC-2014}, a \ac{gp} based local dynamical model allows standard robust control methods to be used on the partially unknown system directly.
Another~\ac{gp} based local dynamical model is proposed in~\cite{IP-Pan-PDDP-2014} to integrate \ac{gp} model with dynamic programming. 
In these two cases, the nonlinear optimization problems considered are \textit{unconstrained}.

In this paper, we propose a different~\ac{gp} based local model.
In this local model,
$\mathbf{x}_k$ in (\ref{eqn:nltvsys}) is replaced by $\mathbf{s}_k= [\meanvalue_k, \textbf{vec}(\sqrt{\varvalue_k})]^T\in \mathbb{R}^{n+n^2}$.
Here, $\textbf{vec}(\cdot)$ denotes the vectorization of a matrix
~\footnote{$\varvalue_k$ is a real symmetric matrix therefore can be diagonalized. 
	The square root of a diagonal matrix can simply be obtained by computing the square roots of diagonal entries.}
Hence (\ref{eqn:nltvsys}) becomes
\begin{equation}
\mathbf{s}_{k+1}=\mathcal{F}'\left(\mathbf{s}_k, \mathbf{u}_k\right) 
\label{eqn:gpdynamicmodel-2}
\end{equation}
Linearizing at the operating point ($\mathbf{s}_k^*, \mathbf{u}^*_k$)
where $\mathbf{s}_k^* = [\meanvalue^*_k,\textbf{vec}(\sqrt{\varvalue^*_k})]^T$, we have
\begin{equation}
\Delta\mathbf{s}_{k+1} =\frac{\partial\mathcal{F}'}{\partial\mathbf{s}_k}\Delta\mathbf{s}_k+\frac{\partial\mathcal{F}'}{\partial\mathbf{u}_k}\Delta\mathbf{u}_k
\label{eqn:gplocalmodel-2}
\end{equation}
Here, $\Delta\mathbf{s}_k = \mathbf{s}_k-\mathbf{s}_k^*$ and $\Delta\mathbf{u}_k = \mathbf{u}_k-\mathbf{u}_k ^*$.
The Jacobian matrices are
\begin{subequations}\label{eqn:extendgpJacobian}
	\begin{align}
	\frac{\partial\mathcal{F}'}{\partial\mathbf{s}_k}&=
	\left[
	\begin{array}{cc}
	\frac{\partial\meanvalue_{k+1}}{\partial\meanvalue_k} & \frac{\partial\meanvalue_{k+1}}{\partial\sqrt{\varvalue_k}}\\
	\frac{\partial\sqrt{\varvalue_{k+1}}}{\partial\meanvalue_k} &
	\frac{\partial\sqrt{\varvalue_{k+1}}}{\partial\sqrt{\varvalue_k}}
	\end{array}
	\right]\in \mathbb{R}^{(n+n^2)\times(n+n^2)} \\
	\frac{\partial\mathcal{F}'}{\partial\mathbf{u}_k}&=
	\left[
	\begin{array}{c}
	\frac{\partial\meanvalue_{k+1}}{\partial\mathbf{u}_k}\\
	\frac{\partial\sqrt{\varvalue_{k+1}}}{\partial\mathbf{u}_k}
	\end{array}
	\right] \in \mathbb{R}^{(n+n^2)\times m}
	\end{align}
\end{subequations}
with the entries given by
\begin{subequations}\label{eqn:computationextendgpJacobian}
	\begin{align}
	\frac{\partial\meanvalue_{k+1}}{\partial\sqrt{\varvalue_k}}&=
	\frac{\partial\meanvalue_{k+1}}{\partial\varvalue_k}
	\frac{\partial\varvalue_k}{\partial\sqrt{\varvalue_k}}\\
	\frac{\partial\sqrt{\varvalue_{k+1}}}{\partial\meanvalue_k}&=
	\frac{\partial\sqrt{\varvalue_{k+1}}}{\partial\varvalue_{k+1}}
	\frac{\partial\varvalue_{k+1}}{\partial\meanvalue_k}\\
	\frac{\partial\sqrt{\varvalue_{k+1}}}{\partial\sqrt{\varvalue_k}}&=
	\frac{\partial\sqrt{\varvalue_{k+1}}}{\partial\varvalue_{k+1}}
	\frac{\partial\varvalue_{k+1}}{\partial\varvalue_k}
	\frac{\partial\varvalue_k}{\partial\sqrt{\varvalue_k}}\\
	\frac{\partial\sqrt{\varvalue_{k+1}}}{\partial\mathbf{u}_k}&=
	\frac{\partial\sqrt{\varvalue_{k+1}}}{\partial\varvalue_{k+1}}
	\frac{\partial\varvalue_{k+1}}{\partial\mathbf{u}_k}
	\end{align}
\end{subequations}
Since $\frac{\partial\sqrt{\varvalue_k}}{\partial\varvalue_k}=\frac{1}{2\sqrt{\varvalue_k}}$ and $\frac{\partial\sqrt{\varvalue_{k+1}}}{\partial\varvalue_{k+1}}=\frac{1}{2\sqrt{\varvalue_{k+1}}}$,
they can be expressed as
\begin{subequations}
	\begin{align}
	\frac{\partial\meanvalue_{k+1}}{\partial\varvalue_k} &=\frac{\partial\meanvalue_{k+1}}{\partial\tilde{\meanvalue}_k}\frac{\partial\tilde{\meanvalue}_k}{\partial\varvalue_k}
	+\frac{\partial\meanvalue_{k+1}}{\partial\tilde{\varvalue}_k}\frac{\partial\tilde{\varvalue}_k}{\partial\varvalue_k}\\
	\frac{\partial\varvalue_{k+1}}{\partial\meanvalue_k} &=\frac{\partial\varvalue_{k+1}}{\partial\tilde{\meanvalue}_k}\frac{\partial\tilde{\meanvalue}_k}{\partial\meanvalue_k}
	+\frac{\partial\varvalue_{k+1}}{\partial\tilde{\varvalue}_k}\frac{\partial\tilde{\varvalue}_k}{\partial\meanvalue_k}\\
	\frac{\partial\varvalue_{k+1}}{\partial\varvalue_k} &=\frac{\partial\varvalue_{k+1}}{\partial\tilde{\meanvalue}_k}\frac{\partial\tilde{\meanvalue}_k}{\partial\varvalue_k}
	+\frac{\partial\varvalue_{k+1}}{\partial\tilde{\varvalue}_k}\frac{\partial\tilde{\varvalue}_k}{\partial\varvalue_k}\\
	\frac{\partial\varvalue_{k+1}}{\partial\mathbf{u}_k} &=\frac{\partial\varvalue_{k+1}}{\partial\tilde{\meanvalue}_k}\frac{\partial\tilde{\meanvalue}_k}{\partial\mathbf{u}_k}
	+\frac{\partial\varvalue_{k+1}}{\partial\tilde{\varvalue}_k}\frac{\partial\tilde{\varvalue}_k}{\partial\mathbf{u}_k}
	\end{align}
\end{subequations}
$\frac{\partial\tilde{\meanvalue}_k}{\partial\varvalue_k}$ and 
$\frac{\partial\tilde{\varvalue}_k}{\partial\varvalue_k}$ can be easily obtained based on (\ref{eqn:augementedstateMeanVar}).
Elaborations of
$\frac{\partial\varvalue_{k+1}}{\partial\tilde{\meanvalue}_k}$ and
$\frac{\partial\varvalue_{k+1}}{\partial\tilde{\varvalue}_k}$
can be found in~\cite{PHD-Deisenroth-EfficientRLusingGP-2010}.

\subsection{Problem Reformulation}
Based on the local model derived above,
define the state variable as
\begin{eqnarray}
\mathbf{Z}_{k+1} & = & \left[ \mathbf{s}_{k+1|k},\cdots, \mathbf{s}_{k+H|k}\right]^T\in\mathbb{R}^{H(n+n^2)} \nonumber\\
&=&[\meanvalue_{k+1},\sqrt{\varvalue_{k+1}},\cdots,\meanvalue_{k+H},\sqrt{\varvalue_{k+H}}]^T
\end{eqnarray}
Also, let
\begin{eqnarray}
\mathbf{U}_k&=& \left[ \mathbf{u}_k,\cdots,\mathbf{u}_{k+H-1} \right]^T\in\mathbb{R}^{Hm}\\
\mathbf{r}_{k+1}^*&=& \left[ \mathbf{r}_{k+1},\mathbf{0},\cdots,\mathbf{r}_{k+H},\mathbf{0}\right]^T\in\mathbb{R}^{H(n+n^2)}
\end{eqnarray}
Problem (\ref{eqn:optproblem2}) then becomes
\begin{subequations}\label{eqn:optproblem3-1}
	\begin{align}
	\min_{\mathbf{U}}&\left\{ \left\|\mathbf{Z}_{k+1}-\mathbf{r}_{k+1}^*\right\|^2_{\tilde{\mathbf{Q}}}+\left\|\mathbf{U}_{k+1}\right\|^2_{\tilde{\mathbf{R}}} \right\}\\
	\mbox{s.t.}\quad
	&\mathbf{I}_{Hn}\mathbf{x}_{\text{min}}\leq\mathbf{M}_{z}\mathbf{Z}_{k+1}\leq\mathbf{I}_{Hn}\mathbf{x}_{\text{max}}\\
	&\mathbf{I}_{Hm}\mathbf{u}_{\text{min}}\leq\mathbf{U}_{k+1}\leq\mathbf{I}_{Hm}\mathbf{u}_{\text{max}}
	\end{align}
\end{subequations}
where
\begin{equation}
\begin{aligned}
\tilde{\mathbf{Q}}=\text{diag}\{[&\mathbf{Q},\text{diag}\{\textbf{vec}(\mathbf{Q})\},\cdots,\mathbf{Q},\\
&\text{diag}\{\textbf{vec}(\mathbf{Q})\}]\}\in\mathbb{R}^{H(n+n^2)\times H(n+n^2)}
\end{aligned}
\end{equation}
$\tilde{\mathbf{R}}=\text{diag}\{[\mathbf{R},\cdots,\mathbf{R}]\}\in\mathbb{R}^{Hm\times Hm}$,
$\mathbf{I}_{a}\in\mathbb{R}^{a}$ is the identity vector,
and
\begin{equation}
\mathbf{M}_z=\left[ 
\begin{array}{cccccc}
\mathbf{I}^T_{n}& 2\mathbf{I}^T_{n^2}& \mathbf{0}&\mathbf{0}&\cdots &\mathbf{0}\\
\mathbf{0}& \mathbf{0}& \mathbf{I}^T_{n}& 2\mathbf{I}^T_{n^2} &\cdots &\mathbf{0}\\
\vdots& \vdots& \vdots& \vdots& \vdots& \vdots\\
\mathbf{0}& \mathbf{0}& \mathbf{0} & \cdots &\mathbf{I}^T_{n}& 2\mathbf{I}^T_{n^2}
\end{array}\right]\in\mathbb{R}^{H\times H(n+n^2)}
\end{equation}
Let $\mathbf{T}_u\in\mathbb{R}^{Hm\times Hm}$ be a lower triangular matrices with unit entries.
Then,
\begin{equation}\label{eqn:linearU} 
\mathbf{U}_k=\mathbf{I}_{Hm}\mathbf{u}_{k-1}+\mathbf{T}_u\Delta\mathbf{U}_k.
\end{equation}
The change in $\mathbf{Z}_{k+1}$ can be expressed as
\begin{equation}
\Delta\mathbf{Z}_{k+1}=\tilde{\mathbf{A}}\Delta\mathbf{s}_{k}+\tilde{\mathbf{B}}\Delta\mathbf{U}_k
\end{equation}
based on the local model, 
with the state and control matrices given by
\begin{subequations}
	\begin{align}
	\tilde{\mathbf{A}}=&\left[\mathbf{A}, \mathbf{A}^2, \cdots, \mathbf{A}^H\right]^T\in\mathbb{R}^{H(n+n^2)}\\
	\tilde{\mathbf{B}}=&\left[\begin{array}{cccc}
	\mathbf{B} &\mathbf{0}  &\cdots &\mathbf{0}\\
	\mathbf{AB}&\mathbf{B}  &\cdots &\mathbf{0}\\
	\vdots     & \vdots & \vdots & \vdots\\
	\mathbf{A}^{H-1}\mathbf{B}&\mathbf{A}^{H-2}\mathbf{B}&\cdots&\mathbf{B}
	\end{array}\right] \in\mathbb{R}^{H(n+n^2)\times Hm}
	\end{align}
\end{subequations}
where $\mathbf{A}$ and $\mathbf{B}$ are the two Jacobian matrices 
(\ref{eqn:extendgpJacobian}) and (\ref{eqn:computationextendgpJacobian}) respectively.
The corresponding state variable $\mathbf{Z}_{k+1}$ is therefore given by
\begin{equation}\label{eqn:linearZ}
\mathbf{Z}_{k+1}=\mathbf{s}_k+\mathbf{T}_z\left(\tilde{\mathbf{A}}\Delta\mathbf{s}_{k}+\tilde{\mathbf{B}}\Delta\mathbf{U}_k\right)
\end{equation}
where $\mathbf{T}_z\in\mathbb{R}^{H(n+n^2)\times H(n+n^2)}$ denotes a lower triangular matrix with unit entries.

Based on (\ref{eqn:linearU}) and (\ref{eqn:linearZ}),
problem (\ref{eqn:optproblem3-1}) can be expressed in a more compact form as
\begin{subequations}\label{eqn:optproblem3-2}
	\begin{align}
	\min_{\Delta\mathbf{U}}&\frac{1}{2}\left\| \Delta\mathbf{U}_k\right\|^2_{\bm{\Phi}}+\bm{\psi}^T\Delta\mathbf{U}_k+\mathbf{C}\\
	\mbox{s.t.} \quad 
	&\label{eqn:optproblem3-2-constraints}\Delta\mathbf{U}_{\text{min}}\leq
	\left[\begin{array}{c}
	\mathbf{T}_u \\\mathbf{T}_z\tilde{\mathbf{B}}
	\end{array}\right]
	\Delta\mathbf{U}_k\leq\Delta\mathbf{U}_{\text{max}}
	\end{align}
\end{subequations}
where
\begin{subequations}\label{eqn:condenseConstraints}
	\begin{align}
	\bm{\Phi}=&\tilde{\mathbf{B}}^T\mathbf{T}_z^T\tilde{\mathbf{Q}}\mathbf{T}_z\tilde{\mathbf{B}}+\mathbf{T}_u^T\tilde{\mathbf{R}}\mathbf{T}_u\in\mathbb{R}^{Hm\times Hm}\\
	\bm{\psi}=&2(\mathbf{s}_k\tilde{\mathbf{Q}}\mathbf{T}_z\tilde{\mathbf{B}}+\Delta\mathbf{s}_k\tilde{\mathbf{A}}^T\tilde{\mathbf{Q}}\tilde{\mathbf{B}}\\
	&-\mathbf{r}_{k+1}^*\tilde{\mathbf{Q}}\mathbf{T}_z\tilde{\mathbf{B}}+\mathbf{u}_{k-1}\tilde{\mathbf{R}}\mathbf{T}_u)\in\mathbb{R}^{Hm}\\
	\mathbf{C}=&(\mathbf{s}_k^2+\mathbf{r}^*_{k+1})\tilde{\mathbf{Q}}+2\mathbf{s}_k\Delta\mathbf{s}_k\tilde{\mathbf{Q}}\mathbf{T}_z\tilde{\mathbf{A}}\\
	&+\mathbf{u}_{k-1}^2\tilde{\mathbf{R}}+\Delta\mathbf{s}_k^2\tilde{\mathbf{A}}^T\tilde{\mathbf{Q}}\tilde{\mathbf{A}}\\
	&-2\mathbf{r}_{k+1}^*(\mathbf{s}_k\tilde{\mathbf{Q}}-\Delta\mathbf{s}_k\tilde{\mathbf{Q}}\mathbf{T}_z\tilde{\mathbf{A}})\\
	\Delta\mathbf{U}_{\text{min}}
	=&\left[\begin{array}{c}
	\mathbf{I}_{Hm}(\mathbf{u}_{\text{min}}-\mathbf{u}_{k-1})\\
	\mathbf{I}_{H(n+n^2)}(\mathbf{x}_{\text{min}}-\mathbf{s}_k-\mathbf{T}_z\tilde{\mathbf{A}}\Delta\mathbf{s}_k)
	\end{array}\right]\\
	\Delta\mathbf{U}_{\text{max}}
	=&\left[\begin{array}{c}
	\mathbf{I}_{Hm}(\mathbf{u}_{\text{max}}-\mathbf{u}_{k-1})\\
	\mathbf{I}_{H(n+n^2)}(\mathbf{x}_{\text{max}}-\mathbf{s}_k-\mathbf{T}_z\tilde{\mathbf{A}}\Delta\mathbf{s}_k)
	\end{array}\right]
	\end{align}
\end{subequations}
Since $\tilde{\mathbf{Q}}, \tilde{\mathbf{R}}, \mathbf{T}_z$ and $\mathbf{T}_u$ are positive definite,
$\bm{\Phi}$ is also positive definite.
Hence (\ref{eqn:optproblem3-2}) is a constrained~\ac{qp} problem and is strictly convex.
The solution will therefore be unique and satisfies the~\ac{kkt} conditions.

\subsection{Optimization Using Active-Set}
\label{subsec:activeset}
The optimization problem (\ref{eqn:optproblem3-2}) can be solved by 
an active-set method~\cite{B-Fletcher-PracticalMethods4Opt-1987}.
It iteratively seeks an active (or working) set of constraints
and solve an equality constrained~\ac{qp} problem until the optimal solution is found.
The advantage of this method is that accurate solutions can still be obtained even when they are ill-conditioned or degenerated.
In addition, it is conceptually simple and easy to implement.
A warm-start technique could also be used
to accelerate the optimization process substantially.

Let $\mathbf{G}=[\mathbf{T}_u,\mathbf{T}_z\tilde{\mathbf{B}}]^T$,
the constraint (\ref{eqn:optproblem3-2-constraints}) 
can be written as
\begin{equation}\label{eqn:inequalityConstraints}
\left[\begin{array}{c}
\mathbf{G}\\
-\mathbf{G}
\end{array}\right]\Delta\mathbf{U}
\leq\left[ \begin{array}{c}
\Delta\mathbf{U}_{\text{max}}\\
-\Delta\mathbf{U}_{\text{min}}
\end{array}
\right]
\end{equation}
Ignoring the constant term $\mathbf{C}$, problem (\ref{eqn:optproblem3-2}) becomes
\begin{subequations}\label{eqn:optproblem3-3}
	\begin{align}
	\min_{\Delta\mathbf{U}}&\frac{1}{2}\left\| \Delta\mathbf{U}_k\right\|^2_{\bm{\Phi}}+\bm{\psi}^T\Delta\mathbf{U}_k\\
	&\mbox{s.t.} \quad
	\label{eqn:optproblem3-3-constraints}\tilde{\mathbf{G}}\Delta\mathbf{U}_k\leq\tilde{\Delta}_\mathbf{U}
	\end{align}
\end{subequations}
where $\tilde{\mathbf{G}}=[\mathbf{G}, -\mathbf{G}]^T\in\mathbb{R}^{2H(m+n+n^2)\times Hm}$ and $\tilde{\Delta}_\mathbf{U}=[\Delta\mathbf{U}_{\text{max}}, -\Delta\mathbf{U}_{\text{min}}]^T\in\mathbb{R}^{2H(m+n+n^2)}$.

Let $\bm{\Pi}_{\Delta\mathbf{U}}$ be the set of feasible points,
and $\mathcal{I}=\{1,\cdots,2H(m+n+n^2)\}$ be the constraint index set.
For a feasible point $\Delta\mathbf{U}_k^*\in\bm{\Pi}_{\Delta\mathbf{U}}$,
the index set for the active set of constraints is defined as
\begin{equation}\label{eqn:activeset}
\mathcal{A}(\Delta\mathbf{U}_k^*)=\{i\subseteq\mathcal{I}|\tilde{\mathbf{G}}_{i}\Delta\mathbf{U}_k^*=\tilde{\Delta}_{\mathbf{U},i}\}
\end{equation}
where $\tilde{\mathbf{G}}_i$ is the $i^{th}$ row of  $\tilde{\mathbf{G}}$
and $\tilde{\Delta}_{\mathbf{U},i}$ is the $i^{th}$ row of the $\tilde{\Delta}_{\mathbf{U}}$.
The inactive set is therefore given by
\begin{equation}\label{eqn:inactiveset}
\begin{aligned}
\mathcal{B}(\Delta\mathbf{U}_k^*)
&=\mathcal{I}\setminus\mathcal{A}(\Delta\mathbf{U}_k^*)\\
&=\{i\subseteq\mathcal{I}|\tilde{\mathbf{G}}_i\Delta\mathbf{U}_k^*<\tilde{\Delta}_{\mathbf{U},i}\}
\end{aligned}
\end{equation}
Given any iteration $j$, 
the working set $\mathcal{W}_k^j$ contains all the equality constraints plus the inequality constraints in the active set.
The following~\ac{qp} problem subject to the equality constraints w.r.t. $\mathcal{W}_k^j$ is considered given the feasible points $\Delta\mathbf{U}_k^j\in\bm{\Pi}_{\Delta\mathbf{U}}$:
\begin{subequations}\label{eqn:optproblem3-4}
	\begin{align}
	&\min_{\bm{\delta}^j}\frac{1}{2}\left\| \Delta\mathbf{U}_k^j+\bm{\delta}^j\right\|^2_{\bm{\Phi}}+\bm{\psi}^T(\Delta\mathbf{U}_k^j+\bm{\delta}^j)\\
	=&\min_{\bm{\delta}^j}\frac{1}{2}\left\|\bm{\delta}^j\right\|^2_{\bm{\Phi}}+(\bm{\psi}+\bm{\Phi}\Delta\mathbf{U}^j_k)^T\bm{\delta}^j\\
	\nonumber&+\underbrace{\frac{1}{2}\left\| \Delta\mathbf{U}_k^j\right\|^2_{\bm{\Phi}}+\bm{\psi}^T\Delta\mathbf{U}_k^j}_{\text{constant}}\\
	&\mbox{s.t.} \quad
	\label{eqn:optproblem3-4-constraints}\tilde{\mathbf{G}}_i(\Delta\mathbf{U}_k^j+\bm{\delta}^j)=\tilde{\Delta}_{\mathbf{U},i}, i\in\mathcal{W}_k^j
	\end{align}
\end{subequations}
This problem can be simplified by ignoring the constant term to:
\begin{subequations}\label{eqn:optproblem3-5}
	\begin{align}
	&\min_{\bm{\delta}^j}\frac{1}{2}\left\|\bm{\delta}^j\right\|^2_{\bm{\Phi}}+(\bm{\psi}+\bm{\Phi}\Delta\mathbf{U}^j_k)^T\bm{\delta}^j\\
	=&\min_{\bm{\delta}^j}\frac{1}{2}{\bm{\delta}^j}^T\bm{\Phi}\bm{\delta}^j+(\bm{\psi}+\bm{\Phi}\Delta\mathbf{U}^j_k)^T\bm{\delta}^j\\
	&\mbox{s.t.} \quad
	\label{eqn:optproblem3-5-constraints}\tilde{\mathbf{G}}_i\bm{\delta}^j=\tilde{\Delta}_{\mathbf{U},i}-\tilde{\mathbf{G}}_i\Delta\mathbf{U}_k^j, i\in\mathcal{W}_k^j
	\end{align}
\end{subequations}

By applying the~\ac{kkt} conditions to problem (\ref{eqn:optproblem3-5}),
we can obtain the following linear equations: 
\begin{equation}\label{eqn:activeset-kkt-2}
\underbrace{\left[\begin{array}{cc}
	\bm{\Phi} & \tilde{\mathbf{G}}_{\mathcal{A}}^T\\
	\tilde{\mathbf{G}}_{\mathcal{A}} & \mathbf{0}
	\end{array}\right]}_{\text{Lagrangian Matrix}}
\left[\begin{array}{c}
\bm{\delta}^j \\ \bm{\lambda}^*_k
\end{array}\right]=
\left[\begin{array}{c}
-\bm{\psi}-\bm{\Phi}\Delta\mathbf{U}^j_k\\ 
\tilde{\Delta}_{\mathbf{U},\mathcal{A}}-\tilde{\mathbf{G}}_{\mathcal{A}}\Delta\mathbf{U}_k^j
\end{array}\right]
\end{equation}
where $\bm{\lambda}_k^*\in\mathbb{R}^{2H(m+n+n^2)}$ denotes the vector of Lagrangian multipliers,
$\tilde{\mathbf{G}}_{\mathcal{A}}\subseteq\tilde{\mathbf{G}}$ and $\tilde{\Delta}_{\mathbf{U},\mathcal{A}}\subset\tilde{\Delta}_\mathbf{U}$ are the weighting matrix and the upper bounds of the constraints w.r.t.  $\mathcal{W}_k^j$.
Let the inverse of Lagrangian matrix be denoted by
\begin{equation}
\left[\begin{array}{cc}
\bm{\Phi} & \tilde{\mathbf{G}}_{\mathcal{A}}^T\\
\tilde{\mathbf{G}}_{\mathcal{A}} & \mathbf{0}
\end{array}\right]^{-1}=
\left[\begin{array}{cc}
\mathbf{L}_1 & \mathbf{L}_2^T\\
\mathbf{L}_2 & \mathbf{L}_3\\
\end{array}\right]
\end{equation}
If this inverse exists, then the solution is given by
\begin{subequations}
	\begin{align}
	\bm{\delta}^j &=-\mathbf{L}_1(\bm{\psi}+\bm{\Phi}\Delta\mathbf{U}^j_k)
	+\mathbf{L}_2^T(\tilde{\Delta}_{\mathbf{U},\mathcal{A}}-\tilde{\mathbf{G}}_{\mathcal{A}}\Delta\mathbf{U}_k^j)\\
	\bm{\lambda}_k^*&=-\mathbf{L}_2(\bm{\psi}+\bm{\Phi}\Delta\mathbf{U}^j_k)
	+\mathbf{L}_3(\tilde{\Delta}_{\mathbf{U},\mathcal{A}}-\tilde{\mathbf{G}}_{\mathcal{A}}\Delta\mathbf{U}_k^j)
	\end{align}
\end{subequations}
where
\begin{subequations}
	\begin{align}
	\mathbf{L}_1&=\bm{\Phi}^{-1}-\bm{\Phi}^{-1}\tilde{\mathbf{G}}^T_\mathcal{A}(\tilde{\mathbf{G}}_\mathcal{A}\bm{\Phi}^{-1}\tilde{\mathbf{G}}^T_\mathcal{A})^{-1}\tilde{\mathbf{G}}_\mathcal{A}\bm{\Phi}^{-1}\\
	\mathbf{L}_2&=\bm{\Phi}^{-1}\tilde{\mathbf{G}}^T_\mathcal{A}(\tilde{\mathbf{G}}_\mathcal{A}\bm{\Phi}^{-1}\\
	\mathbf{L}_3&=-(\tilde{\mathbf{G}}_\mathcal{A}\bm{\Phi}^{-1}\tilde{\mathbf{G}}^T_\mathcal{A})^{-1}
	\end{align}
\end{subequations}

If $\bm{\delta}^j\neq 0$,
then the set of feasible points $\Delta\mathbf{U}_k^j$ fails to minimize problem (\ref{eqn:optproblem3-3}).
In this case, the next set of feasible point is computed for the
next iteration by $\Delta\mathbf{U}_k^{j+1}=\Delta\mathbf{U}_k^j+\kappa^j\bm{\delta}^j$
with step size
\begin{equation}\label{eqn:activeset-step}
\kappa^j=\min\left\lbrace 
1, \min_{i\in\mathcal{B}(\Delta\mathbf{U}_k^j)}\frac{\tilde{\Delta}_{\mathbf{U},i}-\tilde{\mathbf{G}}_i\Delta\mathbf{U}_k^j}{\tilde{\mathbf{G}}_i\bm{\delta}^j}
\right\rbrace 
\end{equation}
If $\kappa^j<1$,
the inequality constraint with index $q=\newargmin_{i\in\mathcal{B}(\Delta\mathbf{U}_k^j)}\frac{\tilde{\Delta}_{\mathbf{U},i}-\tilde{\mathbf{G}}_i\Delta\mathbf{U}_k^j}{\tilde{\mathbf{G}}_i\bm{\delta}^j}$ should be ``activated", giving the working set $\mathcal{W}_k^{j+1}=\mathcal{W}_k^j\cup q$.
Otherwise, we have $\mathcal{W}_k^{j+1}=\mathcal{W}_k^j$.

Alternatively, if the solution gives $\bm{\delta}^j=0$,
then the current feasible points $\Delta\mathbf{U}_k^j$ could be the optimal solution.
This can be verified by checking the Lagrangian multiplier $\lambda_k^*=\min_{i\in\mathcal{W}_k^j\cap\mathcal{I}}\bm{\lambda}_{k,i}^*$.
If $\lambda_k^*\geq0$,
the optimal solution of the (\ref{eqn:optproblem3-3}) at sampling time $k$ is found.
Otherwise, this inequality constraint indexed by $p=\newargmin_{i\in\mathcal{W}_k^j\cap\mathcal{I}}\bm{\lambda}_{k,i}^*$ should be removed from the current working set, giving us $\mathcal{W}_k^{j+1}=\mathcal{W}_k^j\setminus p$.
Algorithm~\ref{alg:activeset} summarizes the active-set algorithm used in the GPMPC.

\IncMargin{1em}
\begin{algorithm}[!t]
	\setstretch{0.1}
	\KwIn{}\\
	\nonl \quad the feasible point $\Delta\mathbf{U}_k^0\in\bm{\Pi}_{\Delta\mathbf{U}}$; \\
	\nonl \quad the working set $\mathcal{W}^0=\mathcal{A}(\Delta\mathbf{U}^0_k)$;\\ 
	\BlankLine
	\For {$j=0,1,2,\cdots$}{
		Compute the $\bm{\delta}^j$ and $\bm{\lambda}_k^*$ by solving the linear equations (\ref{eqn:activeset-kkt-2});\\
		\eIf{$\bm{\delta}^j=0$}
		{$\lambda_k^*=\min_{i\in\mathcal{W}_k^j\cap\mathcal{I}}\bm{\lambda}_{k,i}^*$,\\
			$p=\newargmin_{i\in\mathcal{W}_k^j\cap\mathcal{I}}\bm{\lambda}_{k,i}^*$\\
			\eIf{$\lambda_k^*\geq 0$}{$\Delta\mathbf{U}_k^*=\Delta\mathbf{U}^j_k$;\\
				Stop.}
			{$\mathcal{W}_k^{j+1}=\mathcal{W}_k^j\setminus p$;\\
				$\Delta\mathbf{U}^{j+1}_k=\Delta\mathbf{U}^j_k$;}
		}
		{Compute the step length $\kappa^j$ by (\ref{eqn:activeset-step}),\\
			$q=\newargmin_{i\in\mathcal{B}(\Delta\mathbf{U}_k^j)}\frac{\tilde{\Delta}_{\mathbf{U},i}-\tilde{\mathbf{G}}_i\Delta\mathbf{U}_k^j}{\tilde{\mathbf{G}}_i\bm{\delta}^j}$\\
			\eIf{$\kappa^j<1$}
			{$\Delta\mathbf{U}^{j+1}_k=\Delta\mathbf{U}^j_k+\kappa^j\bm{\delta}^j$;\\ $\mathcal{A}(\Delta\mathbf{U}^{j+1}_k)=\mathcal{A}(\Delta\mathbf{U}^j_k)\cup q$;}
			{$\Delta\mathbf{U}^{j+1}_k=\Delta\mathbf{U}^j_k+\bm{\delta}^j$;\\ $\mathcal{A}(\Delta\mathbf{U}^{j+1}_k)=\mathcal{A}(\Delta\mathbf{U}^j_k)$;}}
	}
	\BlankLine
	\caption{Active-set method for solving the resulting convex optimization problem}
	\label{alg:activeset}
\end{algorithm}
\DecMargin{1em}

\subsection{Implementation Issues}

The key to solving equation (\ref{eqn:activeset-kkt-2}) is the inverse of the Lagrangian matrix.
However, $\tilde{\mathbf{G}}_\mathcal{A}$ is not always full ranked.
Thus the Lagrangian matrix is not always invertible.
This problem can be solved by decomposing $\tilde{\mathbf{G}}_\mathcal{A}$ using QR factorization,
giving us
$\mathbf{G}_{\mathcal{A}}^T=\mathcal{Q}
\left[ \mathcal{R} \;\; \mathbf{0} \right]^{T}$
where $\mathcal{R}\in\mathbb{R}^{m_1\times m_1}$ is an upper triangular matrix with $m_1=\text{rank}(\tilde{\mathbf{G}}_\mathcal{A})$.
$\mathcal{Q}\in\mathbb{R}^{Hm\times Hm}$ is an orthogonal matrix that can be further decomposed to $\mathcal{Q}=\left[\mathcal{Q}_1\;\mathcal{Q}_2\right]$ where $\mathcal{Q}_1\in\mathbb{R}^{Hm\times m_1}$ and $\mathcal{Q}_2\in\mathbb{R}^{Hm\times (Hm-m_1)}$.
Thus, $\mathbf{G}_{\mathcal{A}}^T=\mathcal{Q} =\mathcal{Q}_1\mathcal{R}$ and 
\begin{subequations}
	\begin{align}
	\mathbf{L}_1&=\mathcal{Q}_2(\mathcal{Q}_2^T\bm{\Phi}\mathcal{Q}_2)^{-1}\mathcal{Q}_2^T\\
	\mathbf{L}_2&=\mathcal{Q}_1{\mathcal{R}^{-1}}^T-\mathbf{L}_1\bm{\Phi}\mathcal{Q}_1{\mathcal{R}^{-1}}^T\\
	\mathbf{L}_3&=\mathcal{R}^{-1}\mathcal{Q}_1^T\bm{\Phi}\mathbf{L}_2
	\end{align}
\end{subequations}

The second issue relates to using the appropriate warm-start technique to improve the convergence rate of the active-set method.
For GPMPC,
since the changes in the state between two successive sampling instants are usually quite small,
we can simply use the previous value $\Delta\mathbf{U}_k^*$ at sampling time $k$
as the starting point $\Delta\mathbf{U}^0_{k+1}$ for the next sampling time $k+1$.
This warm-start technique is usually employed in \ac{mpc} optimizations because of its proven effectiveness~\cite{A-Wang-FastMPC-2010}.

\section{Simulation Results}
\label{sec:simresults}

\textcolor{\colourred}{The performance of GPMPC for quadrotor trajectory tracking is evaluated by computer simulations}.
The parameters of translational and rotational subsystems in the numerical quadrotor system are the same as those used in~\cite{A-Alexis-SwitchMPC-QuadrotorUAV-2011}.
\textcolor{\colourred}{All simulations are independently repeated $50$ times on a computer with a $3.40$GHz Intel$\circledR$ Core$^{\text{TM}}$ $2$ Duo CPU with $16$ GB RAM, using Matlab$\circledR$ version $8.1$.
The simulation results presented below are the average values from $50$ independent trials.}

Two non-trivial trajectories are used. 
They are referred to as ``Elliptical" and ``Lorenz" trajectories and are shown as red dotted lines in Figure~\ref{fig:GPMPC_Elliptical_Trajectory} and~\ref{fig:GPMPC_Lorenz_Trajectory}) respectively.
The quadrotor subsystems are subject to external Gaussian white noise with zero mean and unit variance.
\textcolor{\colourred}{The constraints on the control inputs of the translational subsystem are
$0\leq U_1(k)\leq 100, -0.2\leq u_x(k)\leq 0.2, -0.2\leq u_y(k)\leq 0.2$ for the ``Elliptical" trajectory, and
they are $-45\leq u_1(k)\leq 0,-2\leq u_x(k)\leq 2, -2\leq u_y(k)\leq 2$ for the ``Lorenz" trajectory}.
For the rotational subsystem, all observations are scaled to the range $[0.1, 0.9]$. 
The inputs are scaled accordingly.
This is necessary because the numerical range of the original data is very large. 
For example, the unscaled angle $\phi$ lies in the range $[-1.57, 1.57]$ while input $U_4$ lies in the range $[-3.2, 6.2]\times 10^{-8}$.
Using the scaled data leads to much improved training results.

\textcolor{\colourred}{To generate observations for \ac{gp} modelling,
the trajectory tracking tasks are first performed by using the \ac{nmpc} strategy proposed in~\cite{B-Grune-NMPC-2011} 
but without input constraints. For each trajectory,
$189$ observations which consist of inputs, states and outputs are collected for use in \ac{gp} model training.}
\textcolor{\colourred}{The initial state and initial control input are zero.
The weighting matrices $\mathbf{Q}$ and $\mathbf{R}$ are identity matrices. 
Sampling frequency $f_s$ is 1 Hz.}

\subsection{Modelling Results}
\begin{table}[!t]
	\caption{Prediction training and test \ac{mse} values of the \ac{gp} models in the ``Elliptical" tracking problems. (``Trans" denotes the translational subsystem and ``Rotate" represents the rotational subsystem.)}
	\label{tb:trackerrors-1}
	\centering
	\begin{tabular}{c| c| c| c| c}  
		\hline
		& Size & Training & Test &\parbox{1cm}{\centering Average Var}\\
		\hline
		\multirow{4}{*}{``Trans"} & 10 & 2.2485e-4 & 1.4806 & \textbf{1.0231}\\
		\cline{2-5}
		& 50 & 4.1787e-6 & 1.2531 & \textbf{0.1074}\\
		\cline{2-5}
		& 100 & 3.0511e-7 & 1.6733e-6 & \textbf{0.0057}\\
		\cline{2-5}
		& 189 & 1.0132e-7 & 1.0132e-7 & \textbf{2.4843e-4}\\
		\hline 
		\multirow{4}{*}{``Rotate"} & 10 & 2.7443e-6 & 2.3232e-4 & \textbf{2.1150e-4}\\
		\cline{2-5}
		& 50 & 3.0020e-8& 1.0502e-6 & \textbf{1.0853e-4}\\
		\cline{2-5}
		& 100 & 2.8578e-9 & 7.5105e-8 & \textbf{1.0620e-4}\\
		\cline{2-5}
		& 189 & 1.0457e-9 & 1.0457e-9 & \textbf{1.0590e-4}\\
		\hline
	\end{tabular}
\end{table}  
\begin{table}[!t]  
	\caption{Prediction training and test \ac{mse} values of the \ac{gp} models in the ``Lorenz" tracking problems.  (``Trans" denotes the translational subsystem and ``Rotat" represents the rotational subsystem.)}
	\label{tb:trackerrors-2}
	\centering
	\begin{tabular}{c| c| c| c| c}
		\hline
		& Size & Training & Test & \parbox{1cm}{\centering Average Var}\\
		\hline
		\multirow{4}{*}{``Trans"} & 10 & 4.0309e-4 & 2.6872 & \textbf{4.7156}\\
		\cline{2-5}
		& 50 & 1.1986e-4 & 1.1820 & \textbf{1.1696}\\
		\cline{2-5}
		& 100 & 6.5945e-6 & 0.0122 & \textbf{0.0105}\\
		\cline{2-5}
		& 189 & 3.0415e-6 & 3.0415e-6 & \textbf{1.0870e-4}\\
		\hline
		\multirow{4}{*}{``Rotate"} & 10 & 1.0511e-5 & 0.0044 & \textbf{3.1641e-4}\\
		\cline{2-5}
		& 50 & 9.4195e-7 & 4.2896e-5 & \textbf{1.0686e-4}\\
		\cline{2-5}
		& 100 &  3.9616e-8 & 2.7571e-6 & \textbf{1.0607e-4}\\
		\cline{2-5}
		& 189 & 9.2117e-9 & 9.2117e-9 & \textbf{1.0566e-4}\\
		\hline
	\end{tabular}
\end{table}

\textcolor{\colourred}{The first set of results show how well the \ac{gp} models are trained with different sizes of training data.
The full set of $189$ data are used for testing.
As given in Table~\ref{tb:trackerrors-1} and~\ref{tb:trackerrors-2},
the obtained~\ac{gp} models capture the training data well as the training \ac{mse} values are small.
The prediction accuracies reflected by the test~\ac{mse} values show a sudden drop when sufficiently large training sizes are used.
The computational time required for training averages from approximately $1.12$ seconds for a size of $10$ to $4.01$ seconds for a size of $189$}.

\newcommand\tparbox[2]{\protect\parbox[t]{#1}{\protect\raggedright #2}}

\subsection{Control Results}
\begin{figure}[!t]
	\centering
	\subfigure[\tparbox{3.5cm}{Translational Subsystem Controlled Outputs}]{
		\centering
		\label{fig:GPMPC_Elliptical_Trans_Outputs}
		\includegraphics[width=\multifigWidth\linewidth]{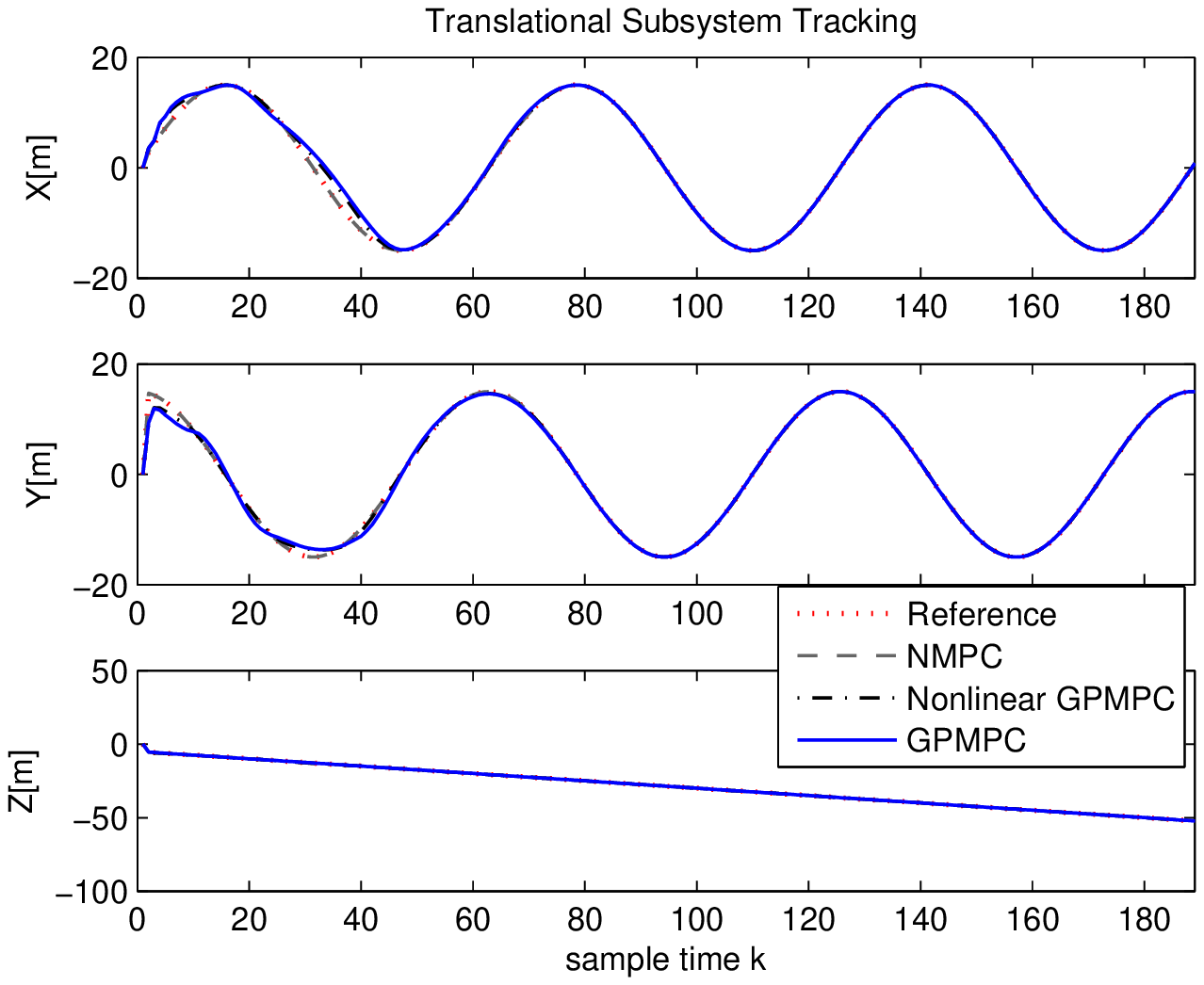}}
	\subfigure[\tparbox{3.5cm}{Translational Subsystem Control Inputs}]{
		\centering
		\label{fig:GPMPC_Elliptical_Trans_Inputs}
		\includegraphics[width=\multifigWidth\linewidth]{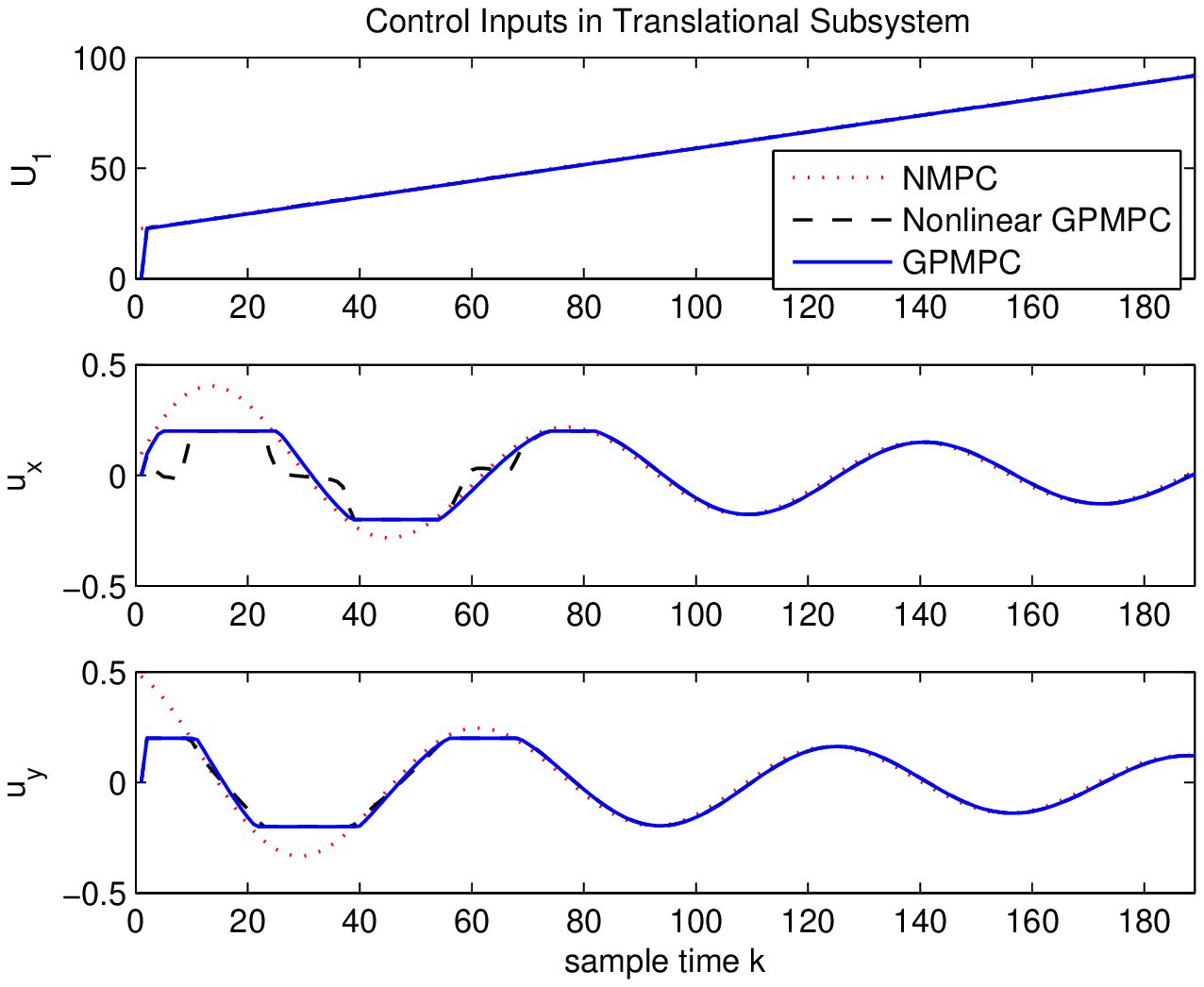}}\\
	\subfigure[\tparbox{3.5cm}{Rotational Subsystem Controlled Outputs}]{
		\centering
		\label{fig:GPMPC_Elliptical_Rotate_Outputs}
		\includegraphics[width=\multifigWidth\linewidth]{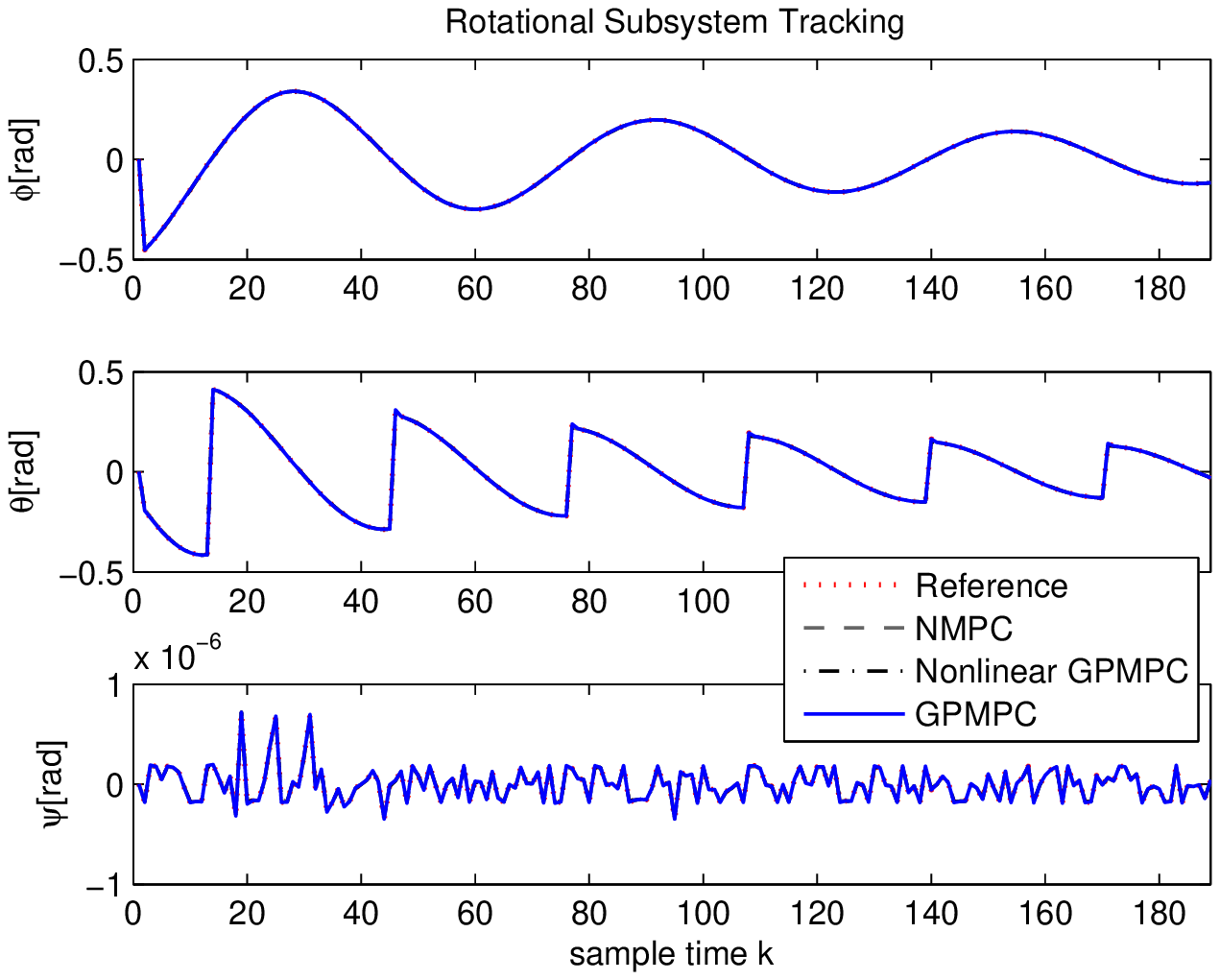}}
	\subfigure[\tparbox{3.5cm}{Rotational Subsystem Control Inputs}]{
		\centering
		\label{fig:GPMPC_Elliptical_Rotate_Inputs}
		\includegraphics[width=\multifigWidth\linewidth]{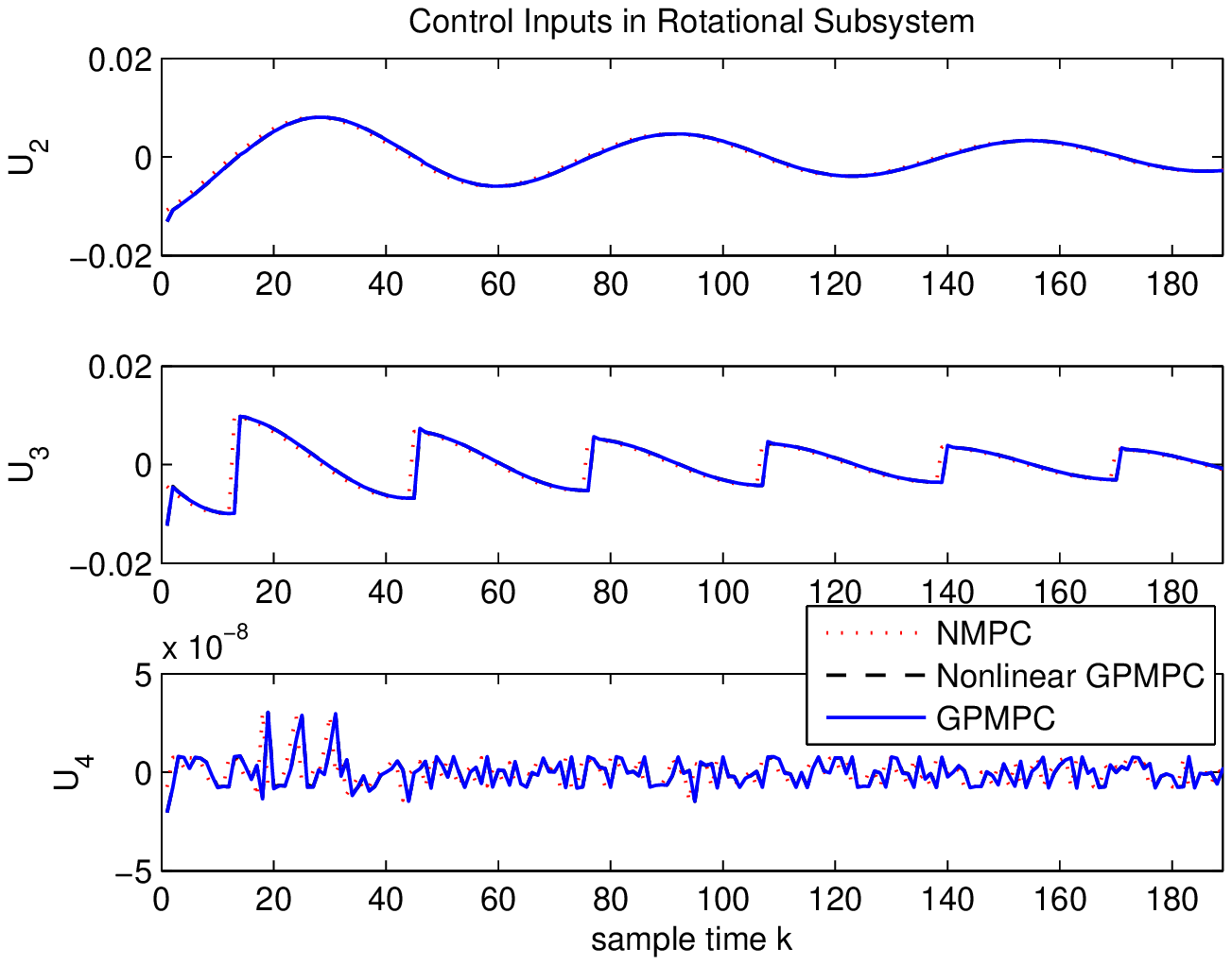}}
	\caption{Controlled outputs and control inputs by using the proposed GPMPC for the both two subsystems in the ``Elliptical" trajectory}
	\label{fig:controlresults_Elliptical}
\end{figure}
\begin{figure}[!t]
	\centering
	\subfigure[\tparbox{3.5cm}{Translational Subsystem Controlled Outputs}]{
		\centering
		\label{fig:GPMPC_Lorenz_Trans_Outputs}
		\includegraphics[width=\multifigWidth\linewidth]{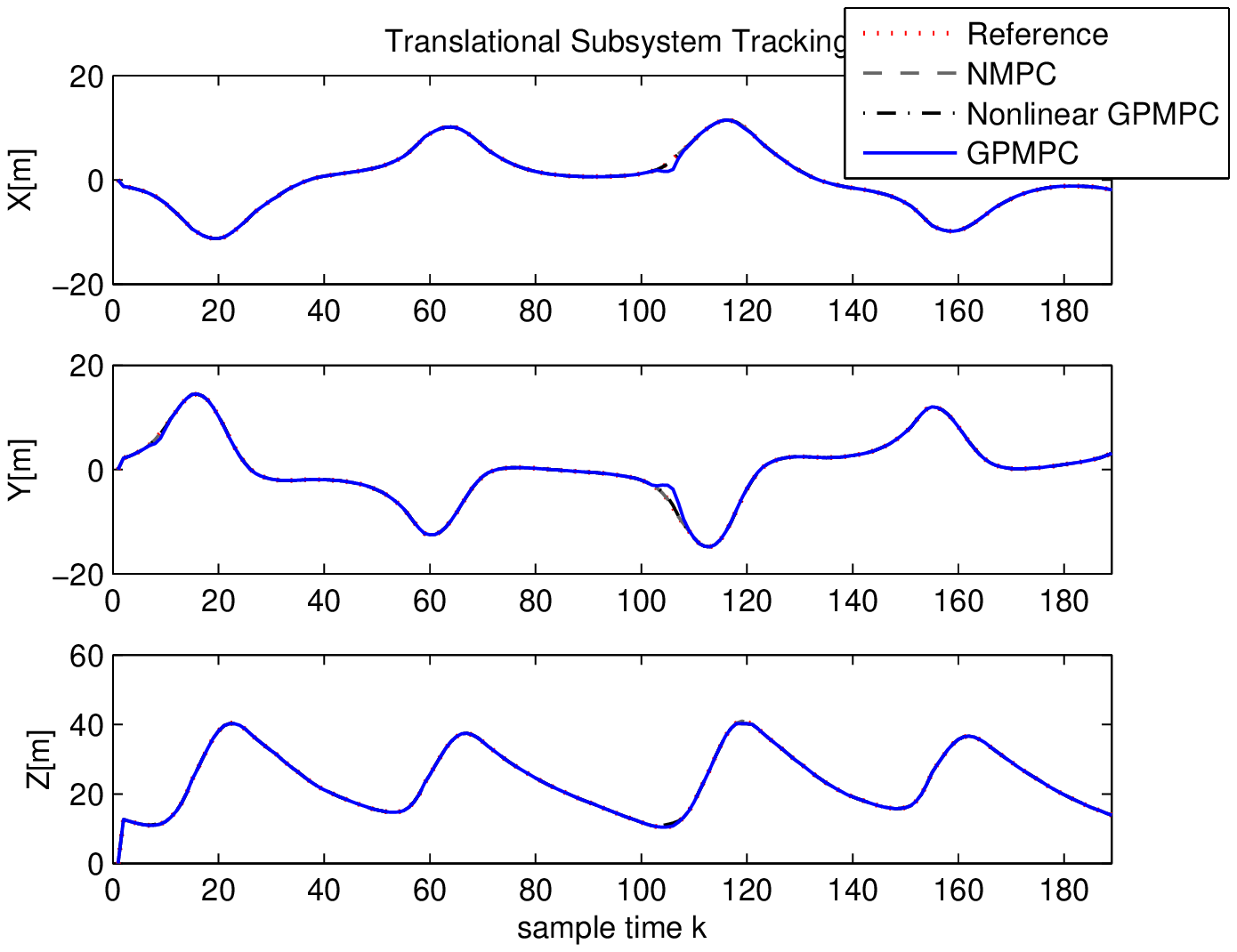}}
	\subfigure[\tparbox{3.5cm}{Translational Subsystem Control Inputs}]{
		\centering
		\label{fig:GPMPC_Lorenz_Trans_Inputs}
		\includegraphics[width=\multifigWidth\linewidth]{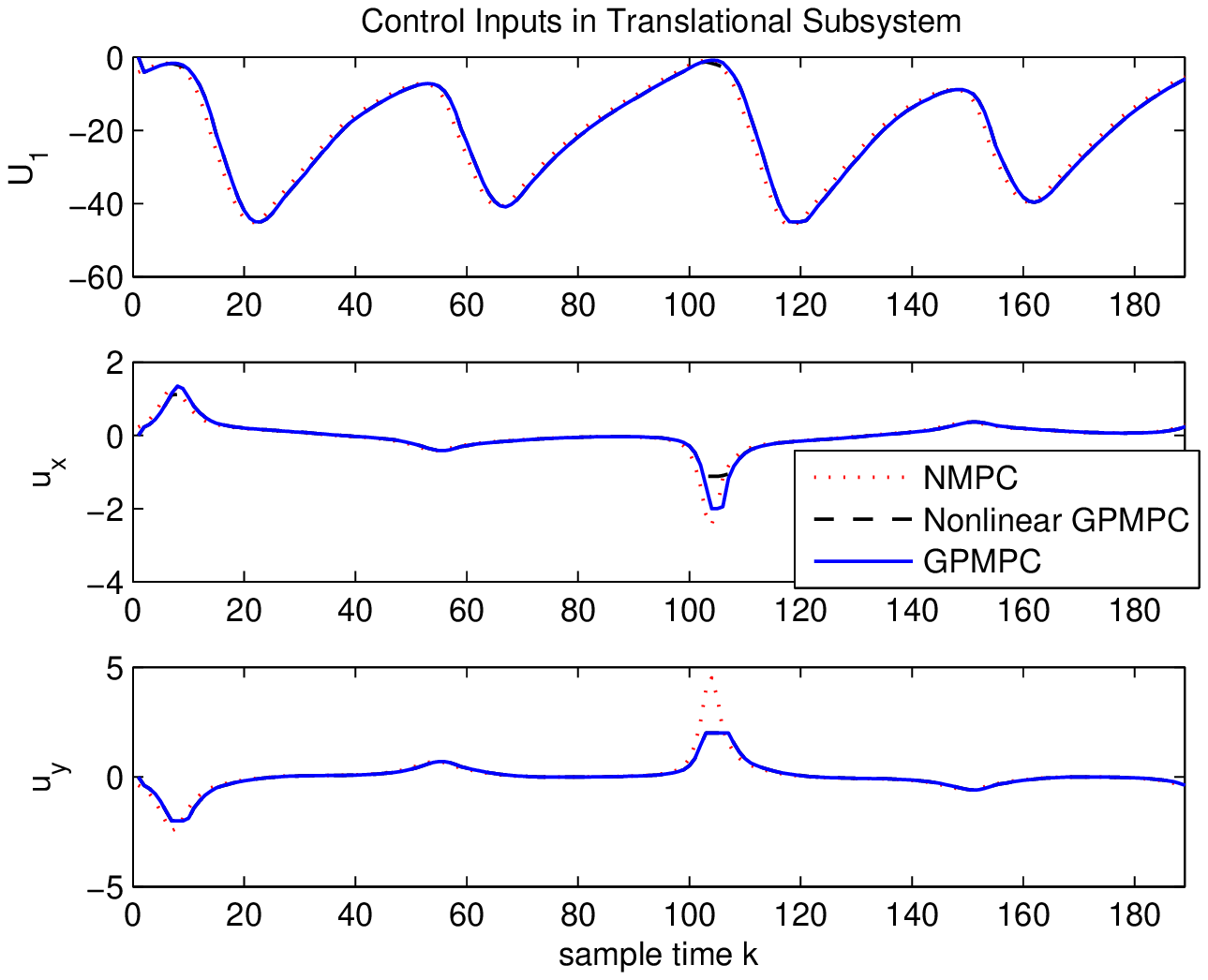}}\\
	\subfigure[\tparbox{3.5cm}{Rotational Subsystem Controlled Outputs}]{
		\centering
		\label{fig:GPMPC_Lorenz_Rotate_Outputs}
		\includegraphics[width=\multifigWidth\linewidth]{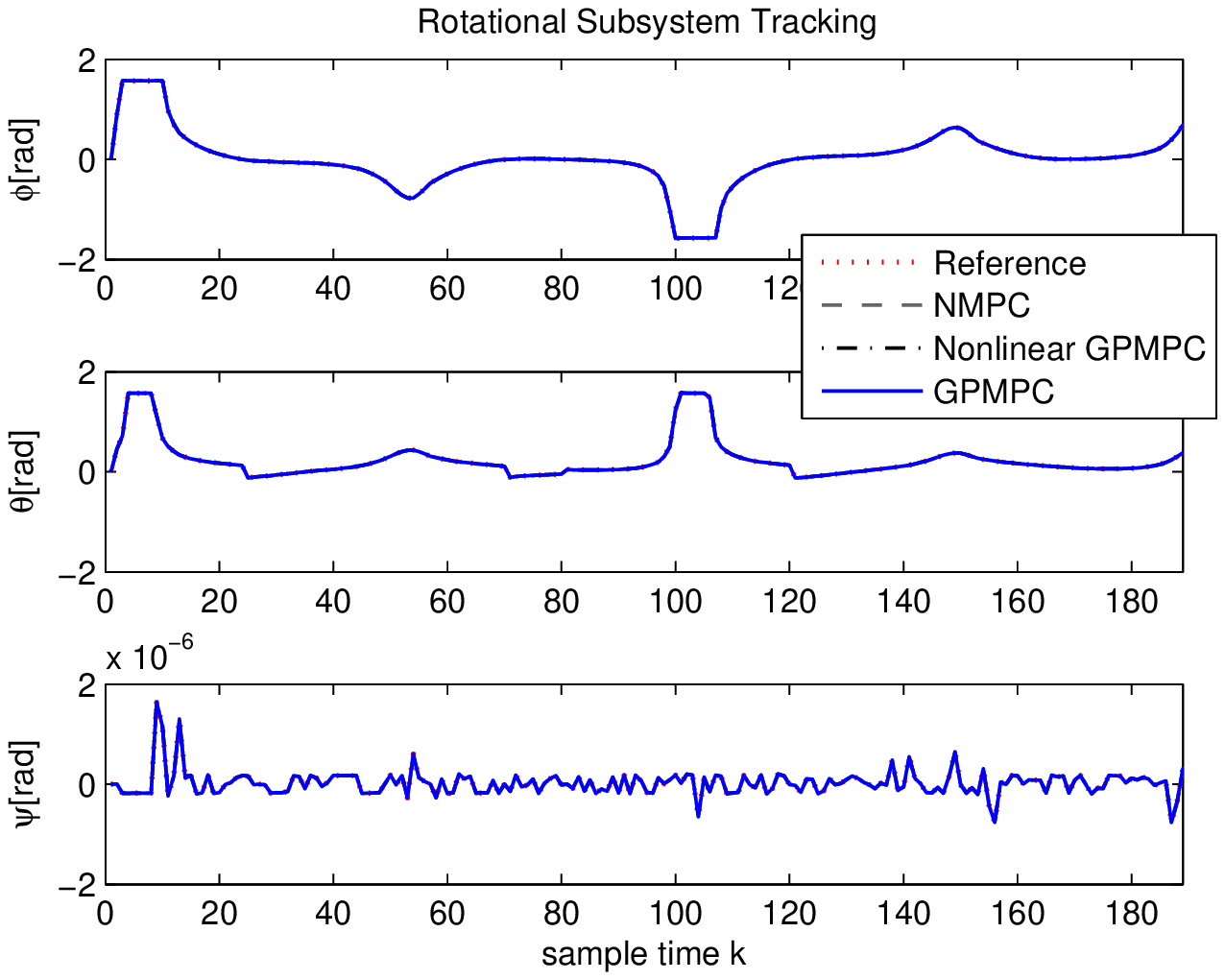}}
	\subfigure[\tparbox{3.5cm}{Rotational Subsystem Control Inputs}]{
		\centering
		\label{fig:GPMPC_Lorenz_Rotate_Inputs}
		\includegraphics[width=\multifigWidth\linewidth]{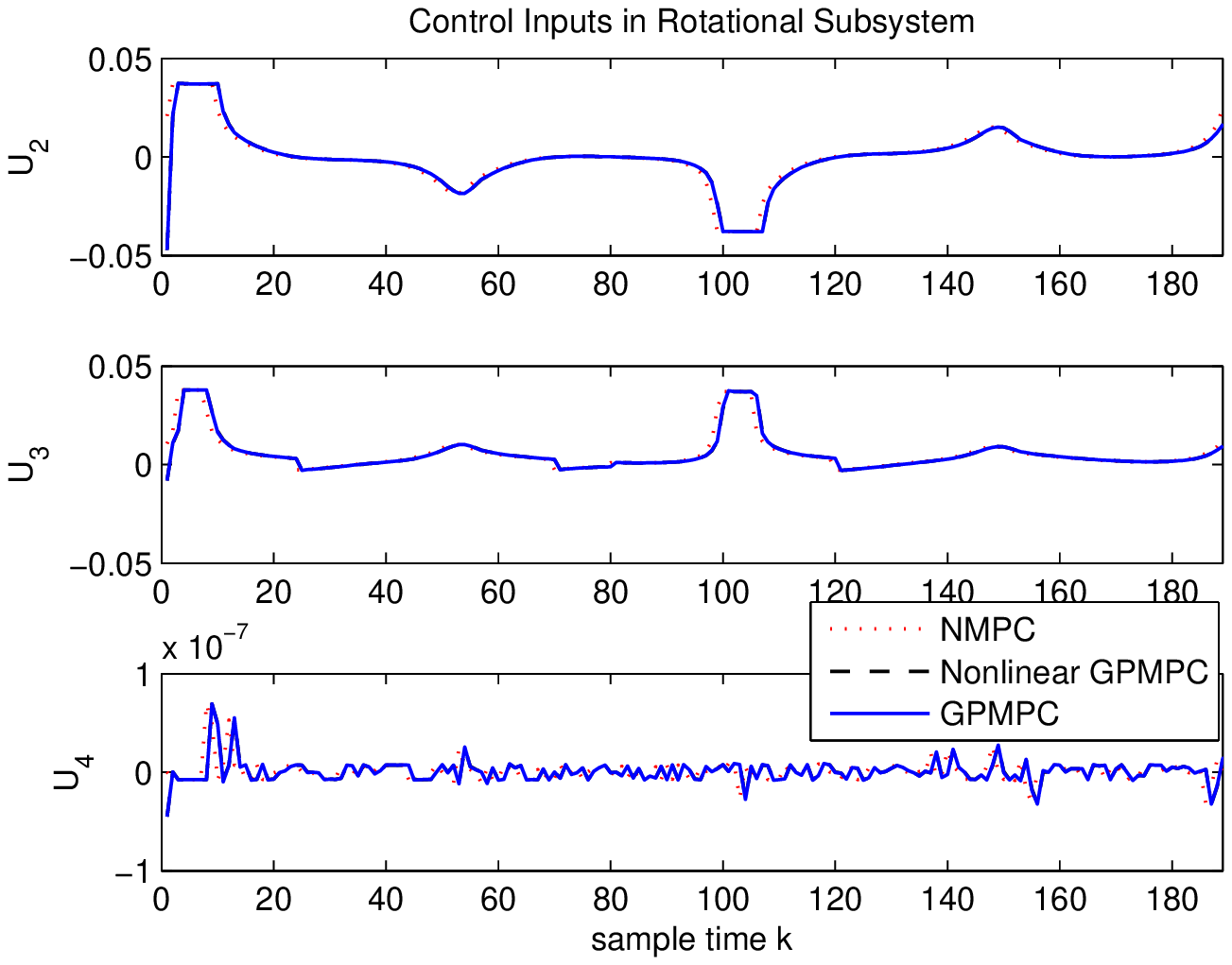}}
	\caption{Controlled outputs and control inputs by using the proposed GPMPC for the both two subsystems in the ``Lorenz" trajectory}
	\label{fig:controlresults_Lorenz}
\end{figure}

\textcolor{\colourred}{The~\ac{gp} models used in the control tasks are trained with all $189$ observations because this
	ensures that the best quality models are obtained.
The performance of using proposed GPMPC scheme is compared with using an exiting~\ac{gp} based~\ac{mpc} algorithm 
(referred to as ``nonlinear GPMPC" or NMPC below) proposed in~\cite{IC-Kocijan-MPCusingGP-2005}.
Even though our optimization problem (\ref{eqn:optproblem2}) with cost function (\ref{eqn:costfunction2-2}) 
is more complicated than the one considered in~\cite{IC-Kocijan-MPCusingGP-2005}, they are essentially similar.	
In addition, we choose $H=1$ as the prediction horizon.
Theoretically, larger values of $H$ is necessary to guarantee the stability of~\ac{mpc} controllers.
However, since solving the nonlinear GPMPC problem with larger values of $H$ effectively is an open problem,
we restrict $H$ to be $1$ in order to make proper comparisons.
Our previous work in~\cite{A-Gang-2016a} has demonstrated that the proposed GPMPC can efficiently be used with a longer horizon.}

\begin{figure}[!t]
	\centering
	\subfigure[``Elliptical"]{
		\centering
		\label{fig:GPMPC_Elliptical_Trajectory}
		\includegraphics[width=\doublefigWidth\linewidth]{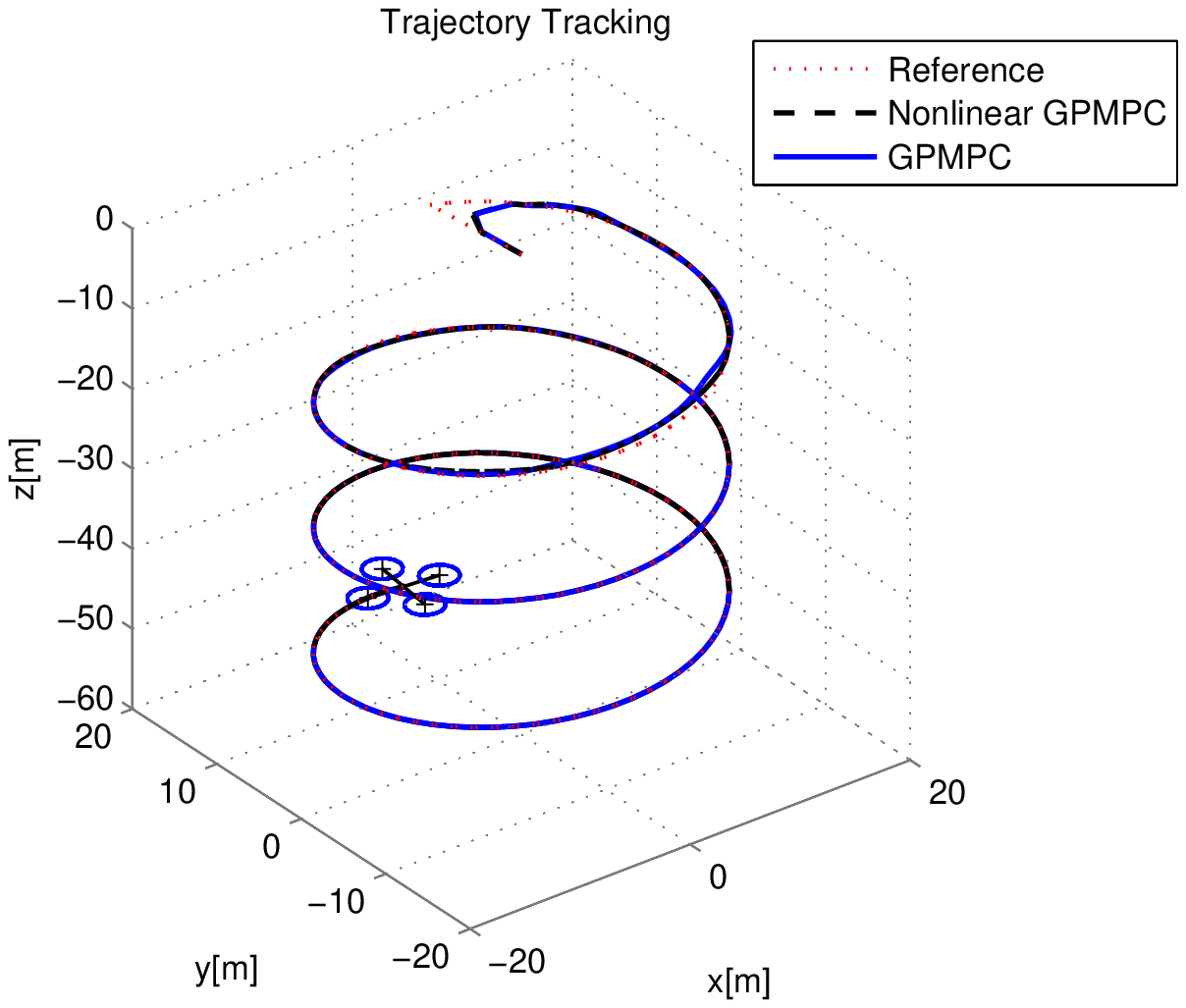}}
	\subfigure[``Lorenz"]{
		\centering
		\label{fig:GPMPC_Lorenz_Trajectory}
		\includegraphics[width=\doublefigWidth\linewidth]{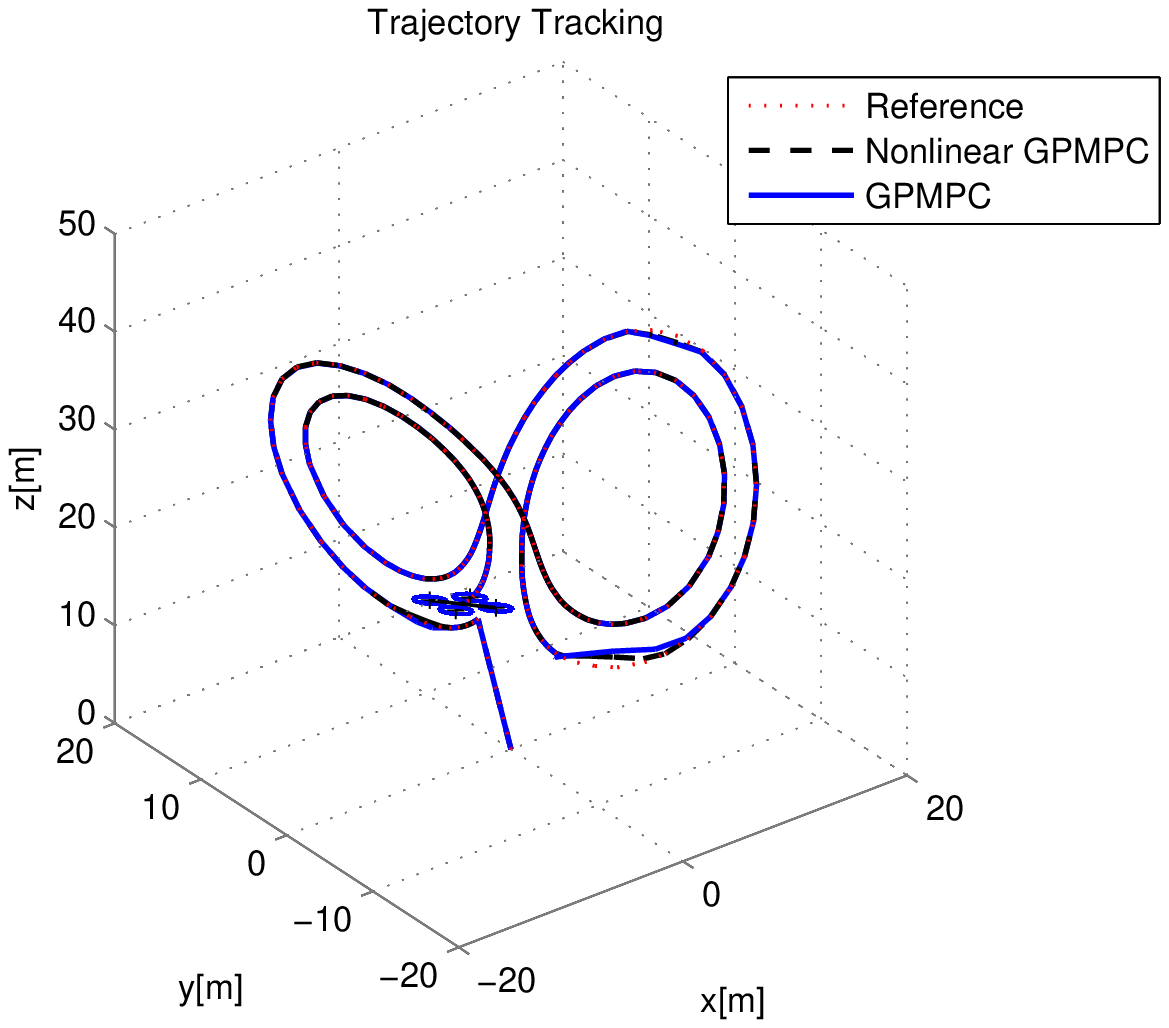}}
	\caption{The results of tracking ``Elliptical" and ``Lorenz" trajectories using the proposed control scheme}
\end{figure}

\textcolor{\colourred}{The control results for the two trajectories 
	are shown in Figures~\ref{fig:controlresults_Elliptical} and~\ref{fig:controlresults_Lorenz}.
They show that NMPC has the best tracking control performance.
However, it should be noted that NMPC does not place any constraints on the control inputs.
In general, the proposed GPMPC is able to closely follow the desired position and attitude values with constrained control inputs.
The overall trajectory tracking results through using the GPMPC based hierarchical control scheme are 
depicted graphically in Figures~\ref{fig:GPMPC_Elliptical_Trajectory} and~\ref{fig:GPMPC_Lorenz_Trajectory}.}

\textcolor{\colourred}{Even with $H=1$, 
	the nonlinear GPMPC requires $220$ seconds and $272$ seconds to compute all $189$ control inputs for the ``Elliptical" and ``Lorenz" trajectory
	respectively.
	This is in contrast to the proposed GPMPC algorithm which only takes $60$ seconds and $56$ seconds.
	This demonstrates that the proposed GPMPC is computationally much more efficient than nonlinear GPMPC}.

\section{Conclusions}
\label{sec:conclude}

\textcolor{\colourred}{A new \ac{mpc} algorithm is proposed for the quadrotor trajectory tracking
	problem where the quadrotor models are trained from empirical data using \ac{gp} techniques.
	hierarchical control scheme based on a computationally efficient 
	\ac{gp} based  for the quadrotor trajectory tracking problem.
Models of the translational and rotational subsystems are learnt from collected data using~\ac{gp} modelling techniques
rather than by traditional Newtonian analysis.
The proposed GPMPC is able to computationally solve the resulting~\ac{mpc} tracking problems which are originally non-convex but can be reformulated as convex ones by using a linearisation of~\ac{gp} models.
The numerical simulation results show that the proposed control scheme is able to closely track non-trivial trajectories.
Its tracking performance is similar to using an \ac{nmpc} method even though GPMPC has input constraints
while no input constraints are placed on \ac{nmpc}.
In addition,
compared to using an existing nonlinear GPMPC,
the proposed GPMPC based control scheme has the advantage of solving the~\ac{mpc} problem much efficiently}.

\begin{appendices}
\section{}
\label{sec:appendix}
\allowdisplaybreaks 

\textcolor{\colourred}{
First rewrite (\ref{eqn:costfunction2-2}) as follows:
\begin{equation}\label{eqn:appendix-cost-1}
\begin{aligned}
&\textit{E}\left[\mathcal{J}(\mathbf{x}_k,\mathbf{u}_{k-1})\right] \\
&=\textit{E}\Big[
\sum_{i=1}^{H}\Big\{\big\|\mathbf{x}_{k+i}-\mathbf{r}_{k+i}\big\|^2_{\mathbf{Q}}+\big\|\mathbf{u}_{k+i-1}\big\|^2_{\mathbf{R}}\} \Big] \\
&=\sum_{i=1}^{H}\bigg\{\underbrace{\textit{E}\Big[ (\mathbf{x}_{k+i}-\mathbf{r}_{k+i})^T\mathbf{Q}(\mathbf{x}_{k+i}-\mathbf{r}_{k+i})\Big]}_{\text{probabilistic term}}\\
& \quad\quad\quad+\underbrace{\mathbf{u}_{k+i-1}^T\mathbf{R}\mathbf{u}_{k+i-1}}_{\text{determinisitc term}}\bigg\}\\
\end{aligned}
\end{equation}}

\textcolor{\colourred}{Let $Q_{ab}$ be the entries of $\mathbf{Q}$ thus $Q_{ab}=[\mathbf{Q}]_{ab}$ and $\varepsilon_{ab}$ as the entries of $\varvalue_k$ thus $\varepsilon_{ab}=[\varvalue_k]_{ab}$,
the ``probabilistic term" can be further derived to
\begin{equation}
\begin{aligned}
&\textit{E}\Big[ (\mathbf{x}_{k+i}-\mathbf{r}_{k+i})^T\mathbf{Q}(\mathbf{x}_{k+i}-\mathbf{r}_{k+i})\Big]\\
&=\textit{E}\Big[\sum_{a=1}^{N}\sum_{b=1}^{N}Q_{ab}(\mathbf{x}_{k+i,a}-\mathbf{r}_{k+i,a})(\mathbf{x}_{k+i,b}-\mathbf{r}_{k+i,b})\Big]\\
&=\sum_{a=1}^{N}\sum_{b=1}^{N}Q_{ab}\textit{E}\Big[(\mathbf{x}_{k+i,a}-\mathbf{r}_{k+i,a})(\mathbf{x}_{k+i,b}-\mathbf{r}_{k+i,b})\Big]\\
&=\sum_{a=1}^{N}\sum_{b=1}^{N}Q_{ab}\bigg\{\textit{E}\big[\mathbf{x}_{k+i,a}-\mathbf{r}_{k+i,a}\big]\textit{E}\big[\mathbf{x}_{k+i,b}-\mathbf{r}_{k+i,b}\big]\\
&\quad+\underbrace{\text{Cov}\bigg((\mathbf{x}_{k+i,a}-\mathbf{r}_{k+i,a}),(\mathbf{x}_{k+i,b}-\mathbf{r}_{k+i,b})\bigg)}_{\varepsilon_{ab}}
\bigg\}\\
&=\sum_{a=1}^{N}\sum_{b=1}^{N}Q_{ab}\Big\{(\meanvalue_{k+i,a}-\mathbf{r}_{k+i,a})(\meanvalue_{k+i,a}-\mathbf{r}_{k+i,a})+\varepsilon_{ab}\Big\}\\
\end{aligned}
\end{equation}
where
\begin{equation}
\begin{aligned}
&\sum_{a=1}^{N}\sum_{b=1}^{N}Q_{ab}(\meanvalue_{k+i,a}-\mathbf{r}_{k+i,a})(\meanvalue_{k+i,a}-\mathbf{r}_{k+i,a})\\ &=(\meanvalue_{k+i}-\mathbf{r}_{k+i})^T\mathbf{Q}(\meanvalue_{k+i}-\mathbf{r}_{k+i})
\end{aligned}
\end{equation}
and
\begin{equation}
\sum_{a=1}^{N}\sum_{b=1}^{N}Q_{ab}\varepsilon_{ab} = \textit{trace}(\mathbf{Q}\varvalue_{k+i})
\end{equation}
Therefore, (\ref{eqn:appendix-cost-1}) can be obtained by
\begin{equation}
\begin{aligned}
&\textit{E}\left[\mathcal{J}(\mathbf{x}_k,\mathbf{u}_{k-1})\right] \\
&=\sum_{i=1}^{H}\bigg\{ (\meanvalue_{k+i}-\mathbf{r}_{k+i})^T\mathbf{Q}(\meanvalue_{k+i}-\mathbf{r}_{k+i})+\mathbf{u}_{k+i-1}^T\mathbf{R}\mathbf{u}_{k+i-1}\\
&\quad+\textit{trace}(\mathbf{Q}\varvalue_{k+i})\bigg\} \\
&=\sum_{i=1}^{H}\Big\{\big\|\meanvalue_{k+i}-\mathbf{r}_{k+i}\big\|^2_{\mathbf{Q}}+\big\|\mathbf{u}_{k+i-1}\big\|^2_{\mathbf{R}}+\textit{trace}\big(\mathbf{Q}\varvalue_{k+i}\big)\Big\}
\end{aligned}
\end{equation}}

\end{appendices}

\end{document}